\documentclass[12pt]{article}

\usepackage{graphicx}
\usepackage{epsfig}
\usepackage{amssymb}
\usepackage{subfigure}
\usepackage{amsmath} 

\evensidemargin - .5truecm
\allowdisplaybreaks[1]

\def \lsim
{\raisebox{-3pt}{$\>\stackrel{<}{\scriptstyle\sim}\>$}}
\def \gsim
{\raisebox{-3pt}{$\>\stackrel{>}{\scriptstyle\sim}\>$}}

\begin{document}

\newcommand{\CC}{{\mathbb C}}
\newcommand{\RR}{{\mathbb R}}
\newcommand{\ZZ}{{\mathbb Z}}
\newcommand{\QQ}{{\mathbb Q}}
\newcommand{\NN}{{\mathbb N}}
\newcommand{\beq}{\begin{equation}}
\newcommand{\eeq}{\end{equation}}
\newcommand{\beal}{\begin{align}}
\newcommand{\eeal}{\end{align}}
\newcommand{\nn}{\nonumber}
\newcommand{\bea}{\begin{eqnarray}}
\newcommand{\eea}{\end{eqnarray}}
\newcommand{\ba}{\begin{array}}
\newcommand{\ea}{\end{array}}
\newcommand{\bfig}{\begin{figure}}
\newcommand{\efig}{\end{figure}}
\newcommand{\bc}{\begin{center}}
\newcommand{\ec}{\end{center}}

\newenvironment{appendletterA}
{
  \typeout{ Starting Appendix \thesection }
  \setcounter{section}{0}
  \setcounter{equation}{0}
  \renewcommand{\theequation}{A\arabic{equation}}
 }{
  \typeout{Appendix done}
 }
\newenvironment{appendletterB}
 {
  \typeout{ Starting Appendix \thesection }
  \setcounter{equation}{0}
  \renewcommand{\theequation}{B\arabic{equation}}
 }{
  \typeout{Appendix done}
 }

%
%
%
%

\begin{titlepage}
\nopagebreak

\renewcommand{\thefootnote}{\fnsymbol{footnote}}
\vskip 2cm
\begin{center}
\boldmath

{\Large\bf Geometry of Winter Model}

\unboldmath
\vskip 1.cm
{\large  U.G.~Aglietti and P.M.~Santini$^*$ }
\vskip .2cm
{\it Dipartimento di Fisica, Universit\`a di Roma ``La Sapienza''  and
\vskip 0.1truecm
 ($\,\,^*$) INFN, Sezione di Roma, I-00185 Rome, Italy} 
\end{center}
\vskip 0.7cm

\begin{abstract}

By constructing the Riemann surface controlling the
resonance structure of Winter model, we determine the 
limitations of perturbation theory.
We then derive explicit non-perturbative results 
for various observables in the weak-coupling regime, 
in which the model has an infinite tower of long-lived 
resonant states.
The problem of constructing proper initial wavefunctions 
coupled to single excitations of the model is also treated  
within perturbative and non-perturbative methods.


\end{abstract}
\vfill
\end{titlepage}    

\setcounter{footnote}{0}

\tableofcontents

\newpage

\section{Introduction}
\label{sect1}

\begin{figure}[ht]
\begin{center}
\includegraphics[width=0.5\textwidth]{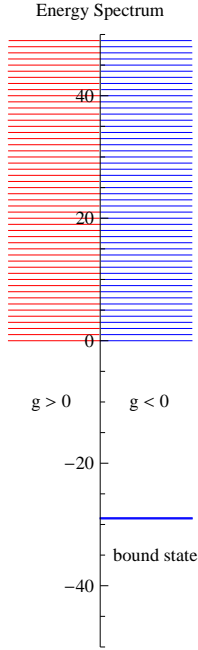}
\footnotesize
\caption{
\label{figura0}
\it Energy spectrum of the Winter model (arbitrary units). 
Repulsive case ($g>0$) on the left of the vertical line (in red); 
attractive case ($g<0$) on the right (in blue). 
Note the discrete line in the latter case.}
\end{center}
\end{figure}

\begin{figure}[ht]
\begin{center}
\includegraphics[width=0.5\textwidth]{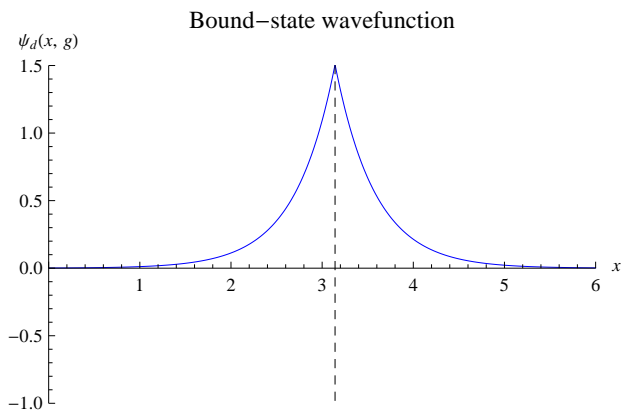}
\footnotesize
\caption{
\label{figura_bs}
\it 
Wavefunction $\psi_d(x,g)$  of the bound state of the Winter model
for $g=-0.07$.  It is also schematically shown the
Dirac potential in $x=\pi$ by means of the dashed vertical line. }
\end{center}
\end{figure}

The only general analytic tool available up to now to study 
``realistic'' quantum field theories, such as for example Quantum
Chromodynamics (QCD) in four dimensions, is perturbation theory.
The great variety of the electromagnetic and hadronic phenomena 
observed at different energies may suggest that exact solutions 
will be beyond human capabilities for a long time, even though there
has been some recent progress in QCD in the $1/N_c$ expansion,
with $N_c $ the number of colors \cite{bochicchio}. 
In general, perturbative cross sections can be written 
schematically as:
\beq
\nonumber
\sigma(g) = \sum_{n=1}^\infty \sigma_n g^n ,
\eeq
where the $\sigma_n$'s are real coefficients and 
$g$ is the coupling of the model.
QCD for instance, because of asymptotic freedom, is weakly coupled 
in the ultraviolet, where perturbation theory is therefore a natural 
tool, while it is strongly coupled in the infrared.
The hadron mass spectrum, the scattering
lengths, the string tension, the chiral symmetry breaking scale, 
the parton distribution functions, the power-corrections to high-energy 
cross sections and event-shape distributions, etc., are all 
well-known examples 
of significant physical quantities which fall outside
the reach of perturbation theory.
Furthermore, being unable to exactly evaluate the $\sigma_n$'s for 
arbitrary $n$, the problem of non-perturbative effects to observables 
is necessarily treated in a rather indirect way.
In the last decades, many different approaches have been developed
for this task: 
methods based on the summation of specific classes of diagrams 
(the $1/N_c$ expansion cited above, the
renormalon calculus supplemented by the large-$\beta_0$ limit, etc.), 
toy models in lower space-time dimensions (typically two or even three),
numerical Monte-Carlo computations of the euclidean theory regularized on
a lattice, direct comparison of perturbative
cross sections with experimental data, and so on.
We may say that a large part of high-energy theoretical activity
of, let's say, the last forty years has been devoted to gain
some control on the non-perturbative effects.
The problem is a recurrent one in QCD and it might have been a problem 
also in the (standard) electroweak theory in the case of a very heavy 
Higgs, let's say $m_H\lsim 800$ GeV. 
In the latter case the scalar sector would become indeed strongly coupled.
However, the Higgs boson was discovered at the Large Hadron Collider (LHC) 
in 2012 with a small mass, 
$m_H\simeq 125$ GeV, as previously indicated by indirect measurements, 
so the problem of non-perturbative electroweak effects has at present 
a limited phenomenological relevance. 
In general, by using a (truncated) perturbative computation 
of a quantum field theory model to describe
some specific phenomenology ranging, let's say, from high-$T_c$ 
superconductivity to strong interactions, one is always faced with
the problem of "what is missing'', i.e. which effects lie in the 
unevaluated terms or are actually "invisible'' to perturbation methods.
In such a situation, it may be interesting to study
a model which, though not a quantum field, can be analyzed
both in perturbative and non-perturbative way. 
It is in this spirit that we present a systematic study of the
so-called Winter model \cite{flugge,winter,mailaggr,primo,secondo}, 
a non-relativistic quantum mechanics model possessing, in the weak-coupling 
regime, an infinite tower of resonant states coupled to a continuum  \cite{gamow}, 
with Hamiltonian in proper units:
\beq
\hat{H} = -\frac{\partial^2}{\partial x^2} + \frac{1}{\pi g} \delta(x-\pi)
\eeq
on the half line $x\ge 0$ with vanishing boundary conditions in the origin,
$\psi(x=0,t)=0$. $g\in\RR$ is the only coupling of the model.
Winter's model has an unstable energy spectrum in the free limit
$g\to 0$: for $g\to 0^+$ the spectrum is uniformly bounded from 
below by zero, while for $g\to 0^-$ the spectrum is unbounded 
because of the appearance of an eigenfunction in the 
discrete spectrum with energy 
$\varepsilon(g) \simeq -1/(4\pi^2 g^2)\to-\infty$
(see fig.\ref{figura0}).
The free limit is therefore a singular one. 
In classical terms, the particle falls in the potential
trap located at $x=\pi$.
The instability is related to the behavior just
of the fundamental state (i.e. of the state with the lowest energy) 
for $g\to 0$, i.e. to the non-trivial ``vacuum structure'' 
in a small neighborhood of the free theory (see fig.\ref{figura_bs}). 
Furthermore, transition amplitudes for $-1<g<0$ contain contributions
of the form $e^{1/g}$, coming from the bound state, non-analytic in the origin, 
which are typical of non-perturbative quantum field theory effects.
A similar instability in Quantum Electrodynamics (QED) 
was conjectured in the 50's by F. Dyson \cite{dyson}
\footnote{
An instability mechanism similar to the QED one also occurs in the scalar
$\lambda \varphi^4$ theory, as well as in the simple case of a (quantum)
anharmonic oscillator \cite{zinnjustin}.
}.
For $0 < \alpha \ll 1$, where $\alpha \equiv e^2/(4\pi)$ 
is the fine structure constant of QED\footnote{
Experimentally $\alpha\cong 1/137$ at low energies.}
with $e$ the electron charge, the fundamental state is the vacuum, 
i.e. the "empty'' state, without any electron-positron pair or any photon.
Because of field fluctuations, electron-positron pairs
(as well as photons) come out of the vacuum as virtual particles 
only, as their creation as real particles would increase the energy.
On the other hand, for $\alpha<0$ particles with equal charges
attract each other, while electrons and positrons repel each other, 
so that the Coulomb energy associated with an $e^+ e^-$ pair 
is positive.
The rest energy of a system containing $N$ pairs is $2 N m c^2$,
with $m$ the electron mass --- proportional to $N$ --- while the 
potential energy is of order $|\alpha| N^2/\lambda_c$ 
--- proportional to $N^2$ --- where $\lambda_C \equiv \hbar/(m_e c)$ 
is the electron Compton wavelength.
That implies that the potential energy overcomes the rest energy for 
large enough $N$. 
As a consequence, a huge quantity of $e^+e^-$ pairs can be created as real 
particles out of the vacuum, which would then decay into a state full of 
electrons and positrons, by lowering its energy to arbitrarily negative 
values. 
The physical picture is the following: the $e^+e^-$ pairs are created close 
to each other, let's say within their Compton wavelength 
(the $QED$ interaction is local) and then,
to minimize the energy, all the electrons fly on
one side, coming close to each other, with the positrons flying 
on the opposite side.
This spontaneous polarization of the vacuum should also occur in the 
usual particle states of the Fock space\footnote{
The above argument is qualitative and does not provide any estimate of
the decay time.}.
This instability of $QED$ in a neighborhood of $\alpha=0$
was related by Dyson to the divergence of the perturbative series in 
$\alpha$.
The spectrum instability of Winter model in a neighborhood of the 
free theory might suggest that also its perturbative expansion around 
$g=0$ is divergent --- perhaps asymptotic to the exact theory for $g\to 0$, 
as is supposed to be in QED. The reality is actually more intricate;
after all, physical intuition is based on $g\in\RR$, while a full
understanding of the model, as we are going to show in detail, requires to 
complexify $g$.
It turns out that the resonances of the model, which behave in space-time
for $x>\pi$ (i.e. outside the cavity) as
\beq
e^{ \, i k x \, - \, i k^2 t}
\eeq
with $k\in\CC$, are controlled by a multivalued function 
\beq
w \, = \, h(z)
\eeq
which is the inverse of the entire function
\beq
z \, = \, \frac{ e^{ w } - 1 }{ w } ,
\eeq 
where $w \equiv 2 \pi i k$ and $z \equiv - g$.\footnote{
The minus sign in front of $g$ is inserted just for practical 
convenience.}
The transcendental (infinite order) multivaluedness of $h$ is related 
to the fact that the model has an infinite tower of resonances.
Each resonance, let's say the $n$-th one with $n$ a non-zero integer,
is associated to a sheet $S_n$ of the Riemann surface $S$ of $h$;
there is also an additional sheet, $S_0$, related to the bound state.
Each $S_n$, with $n\ne 0$, has order-one (square-root) branch points in 
\beq
c_n \, \approx \, \frac{ i }{ 2 \pi n }
\eeq
and at infinity, connecting $S_n$ to $S_0$.
That implies that $S_0$ has a countable set of order-one branch points
$\{c_n\}_{n\ne 0}$ accumulating at the origin as
\beq
\label{accumulateto0}
c_n \to 0 \,\,\,\,\,\,\,\,\,\, {\rm for } \,\,\,\,\, n \to \pm \infty .
\eeq
It also follows that different sheets, $S_n$ and $S_k$ with $n,k\ne 0$, 
"talk to each other" only indirectly, through $S_0$, which is a different 
(and more complicated) sheet with respect to all the other ones.
Let us also observe that the square-root branch point is, in some sense, 
the most general coupling between different sheets one could think of.
The perturbative expansions (series in powers of $g$) of quantities related 
to the $n$-th resonance
(wave-vector, frequency, width, etc.) are convergent within a disk centered 
in $g=0$ of radius 
\beq
R_n \, = \, \left|c_n\right| \approx \, \frac{ 1 }{ 2 \pi n } .
\eeq
The convergence radius is then controlled by the branch point $c_n$ 
connecting $S_n$ to $S_0$.
A non-zero radius of convergence for the expansion
around zero was expected on physical ground but, remaining in 
the physical domain $g\in\RR$, one could not have derived in a natural 
way its value.
The instability of the spectrum, discussed above on physical ground, manifests 
itself mathematically in the fact that, according to eq.(\ref{accumulateto0}), 
the origin of the ``bound-state sheet'' $S_0$ is a non-isolated singularity.
The above property is related to the specific geometric structure of $S$:
if the square-root branch points had coupled, for example, $S_n$ to $S_{n+1}$
for any integer $n$, the point $g=0$ would have been instead an analyticity 
point for $S_0$.
For quantities related to the bound state (such as its energy, 
its wavefunction, etc.) 
no convergent power-series expansion around $g=0$ exists.
While in the case of $QED$ the instability of the vacuum is expected to 
manifest also in any state of the Fock space, in the Winter model it is
restricted to the fundamental state. 
The second implication of eq.(\ref{accumulateto0}) is that 
\beq
R_n \to 0 \,\,\,\,\,\,\,\,\, {\rm for } \,\,\,\,\, n \to \pm \infty.
\eeq
The consequence is that perturbation theory can accurately describe 
the dynamical properties involving a finite number of resonances
for small enough coupling,
but it cannot describe quantities involving an infinite number of them.
We will see that also the second class of observables contains fundamental 
physical quantities. 

Our motivation to further investigate Winter model might suggest that
it is just a toy model for quantum field theory: that is not actually
the case \cite{segre}.
Since it describes particles confined by potential barriers,
with tunable efficiency,
it is still currently used in quantum chemistry, 
together with its natural generalizations 
\cite{refbase,refbase2,hatano}. 
Let's schematically summarize the history of Winter model
relevant to the present work.
As far as we know, this model was originally 
introduced in  \cite{flugge}, where the resonance
properties of the spectrum for a small 
positive coupling were analyzed.
The temporal evolution of metastable states 
of sinusoidal shape concentrated at $t=0$ inside the 
cavity --- the segment $[0,\pi]$ --- was studied in  \cite{winter}.
In this work, post-exponential power corrections 
in time were explicitly calculated, confirming a 
previous general analyticity argument about the breakdown 
of the exponential behavior at very large times.
In ref.\cite{primo} Winter's computation was repeated, finding
additional "non-diagonal" contributions to the exponential 
time evolution, which had been overlooked in \cite{winter}.
These new terms imply a coupling of the initial state
with all the resonances of the model and not just with a
single one. The off-diagonal terms a small coupling $\mathcal{O}(g)$
compared to the one in \cite{winter}, but decay in general
slower in time, dominating then the wavefunction in a large 
temporal region. 
Actually, most of the non-trivial properties of Winter model
--- resonance mixing in particular --- are consequences of this
additional contributions to time evolution recently discovered. 
In \cite{primo} it was also found that the coupling of the initial
state to the resonances was controlled at first order in $g$
by an infinite matrix of the form
\beq
U(g) = 1 + g A,
\eeq
with $A$ a real antisymmetric matrix (see eq.(\ref{Aexplicit})). 
$U(g)$ is then an infinitesimal
rotation in the infinite-dimensional vector space of the resonances.
It was then natural to conjecture that higher-orders in $g$ would
have led to the exponentiated form, i.e. to the unitary matrix
\beq
U(g) = \exp(g A) = 1 + g A + \frac{g^2}{2} A^2 + \cdots.
\eeq   
In order to check this structure, the second-order computation in $g$
of $U(g)$ was made in \cite{secondo}. In addition to the expected term
$g^2 A^2/2$, it was also found a "large" term not compatible with
any generalized form of exponentiation, whose physical interpretation 
was problematic. A possible investigation at that point could have been
to push the perturbative expansion to $\mathcal{O}(g^3)$, in order to have
some hint of the general structure and eventually resum the expansion at all
orders in $g$, in the spirit of classical quantum-field-theory investigations. 
In this work however we follow a different route: we abandon perturbation 
theory and perform an analytic study in the complex plane of the functions 
controlling the expansion above for $U(g)$. As we are going to show, that 
allows us to determine the convergence region of the perturbative expansion
and to find explicit non-perturbative formulas for relevant observables
which replace the perturbative ones outside their convergence region.  

This work is at the border of different areas: 1) quantum field theory,
in particular high-energy theoretical physics, where the interplay
between perturbative and non-perturbative effects is crucial, as 
already discussed; 2) mathematical physics, as we investigate the 
analytic and geometric structure of the Winter model and finally  
3) quantum physics ---quantum chemistry in particular --- where generalized 
Winter models are currently investigated.
Therefore, since the paper is of potential interest to readers with 
different backgrounds, we tried to be as simple and explicit as 
we could. 
The paper is organized as follows.
In sec.\ref{sect2} we summarize the properties of the spectrum of the
Winter model and we discuss in particular the free limit $g\to 0$,
in which the system decomposes in two non-interacting subsystems,
a particle in the box $[0,\pi]$ and a particle in the half-line
$[\pi,\infty)$. All that is in complete agreement with physical 
intuition. 
In sec.\ref{sect3} we schematically discuss the time evolution of 
wavefunctions initially  concentrated inside the cavity and 
of sinusoidal shape. We also treat, in general terms, perhaps the most 
significant effect in the evolution of metastable 
states of Winter model: the above mentioned mixing of the resonances.
In sec.\ref{sect4} we specify the discussion on resonance mixing
by using explicit power expansions in $g$ and we discuss
the problems associated with the perturbative expansion.
In sec.\ref{sect5} we abandon perturbation theory and 
investigate the structure of the Riemann surface
$S$ of the multivalued function $w=h(z)$ 
controlling the behavior of the resonances and
the form of the infinite mixing matrix $U(g)$.
In sec.\ref{sect6} we re-analyze the properties  
of the resonances and of the mixing matrix by means
of the exact (non-perturbative) information obtained
with the previous geometric study.
We deal again, in particular, with the problems encountered 
with perturbation methods.
Finally, in sec.\ref{sect7} we draw our conclusions and 
we discuss some natural developments of our work.
There are also three appendices.
In appendix \ref{appA} we discuss the connection of our function
$h$ with the Lambert $W$ function \cite{knuth}, a kind of generalization
of the complex logarithm. 
In appendix \ref{appB} we present a general treatment, as well as
explicit formulas, for the branch points of the function $h$,
which are crucial in our analysis.
Finally, in appendix \ref{appC} we show that the expansion for large
$|z|$ of $h(z)$, which involves the composition of the complex logarithm
with itself, actually has exactly the same multivaluedness as $h$,
which is just logarithmic.
to
\section{Winter Model}
\label{sect2}

In general, the Hamiltonian operator of the Winter model reads:
\beq
\label{WinterH}
\hat{H} \, = \, - \, \frac{ \hbar^2 }{2 m} \frac{\partial^2}{\partial x^2} 
\, + \, \lambda \, \delta(x - L) \, ,
\eeq
where $m$ is the mass of the particle, $\lambda$ is a (real) coupling constant
and $\delta(x-L)$ is the Dirac $\delta$-function with support in $x=L>0$
\cite{jackiw}.
The domain is the half-line $0\le x < \infty$ and we assume vanishing boundary 
conditions at zero:\footnote{
Equivalently, one may think to the problem in the whole real axis, $x\in\RR$,
with an additional infinite potential for $x<0$.
}
\beq
\psi(x=0,t) = 0 , \, \, \, \, \, \, \, \, \, t \in \RR.
\eeq
Formulas can be simplified by going to a proper adimensional coordinate via
\beq
x = \frac{L} {\pi} x'
\eeq
and rescaling the Hamiltonian as:
\beq
\hat{H} =  \frac{ \hbar^2 \pi^2 }{ 2 m L^2 } \hat{H}'.
\eeq
The new (adimensional) Hamiltonian then takes the form in which it
appears in the introduction:
\beq
\label{eqmod}
\hat{H'} \, = \, - \, \frac{\partial^2}{\partial x'^{\, 2} } \, + \, \frac{1}{\pi g}  \, \delta(x' - \pi),
\eeq
and contains the single (real) parameter 
\beq
g \, = \, \frac{ \hbar^2  }{ 2 m \lambda L } .
\eeq
The time-dependent Schrodinger equation
\beq
i \hbar \frac{\partial \psi }{\partial t }  = \hat{H} \psi  
\eeq
now reads
\beq
i \frac{\partial \psi }{\partial t' }  = \hat{H'} \psi ,
\eeq
where time is rescaled as
\beq
t' \equiv  \frac{ \hbar \pi^2 }{ 2 m L^2 } t .
\eeq
Let us omit primes from now on for simplicity's sake.
It is possible to rescale the Winter Hamiltonian (\ref{WinterH}) in 
different ways, as made for example by Winter itself in 
\cite{winter}.
The main point however is that we deal in any case with a one-parameter model
describing the coupling of a cavity 
(the segment $0\le x \le \pi$) with the outside (the half-line $x\ge\pi$).
As we are going explicitly to show in the next sections and as anticipated 
in the introduction, for $|g|\ll 1$
resonant long-lived states inside the cavity come into play.

\subsection{Spectrum}

\begin{figure}[ht]
\begin{center}
\includegraphics[width=0.5\textwidth]{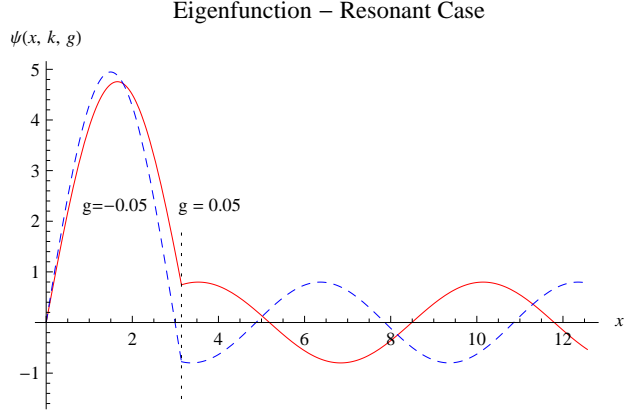}
\footnotesize
\caption{
\label{figura01}
\it Eigenfunction $\psi(x;k,g)$ for resonant $k= 1-g$. 
Repulsive case ($g=0.05$): continuous (red) line; 
Attractive case ($g=-0.05$): dashed (blue) line. 
The enhancement of the amplitude inside the cavity is
clearly visible, as well as the $\mathcal{O}(\psi(x=\pi)/g)$ 
discontinuity of $\psi'(x=\pi)$.
}
\end{center}
\end{figure}

\begin{figure}[ht]
\begin{center}
\includegraphics[width=0.5\textwidth]{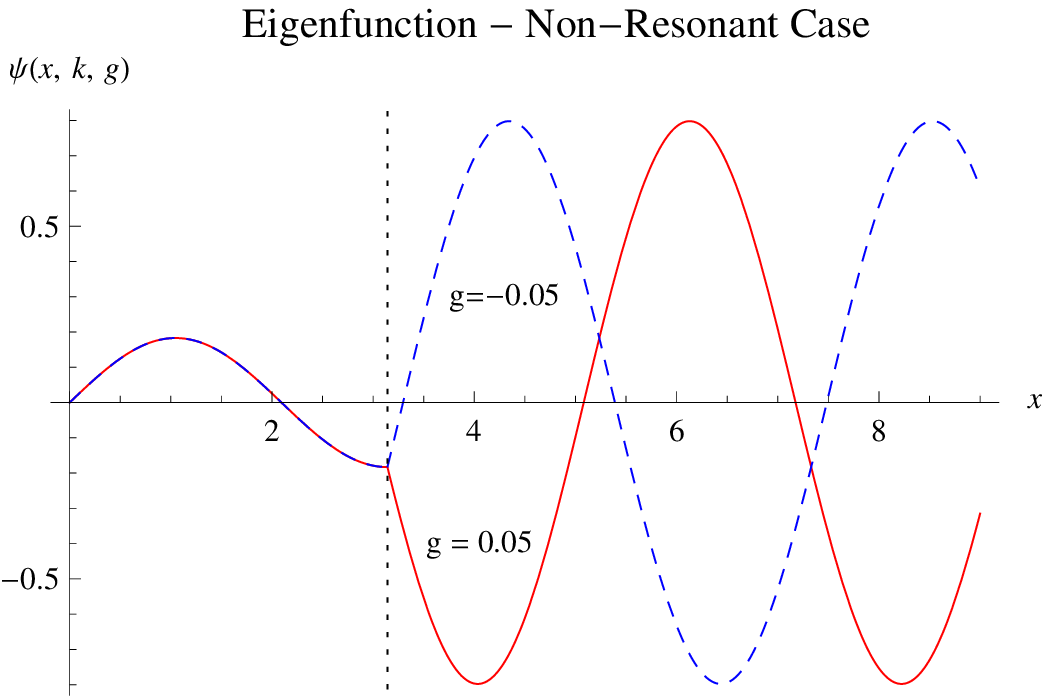}
\footnotesize
\caption{
\label{figura02}
\it Eigenfunction $\psi(x;k,g)$ for non-resonant $k=1.5$. 
Repulsive case ($g=0.05$): continuous (red) line; Attractive case 
($g=-0.05$): dashed (blue) line. The amplitude inside the cavity is much
smaller than outside it.}
\end{center}
\end{figure}

\begin{figure}[ht]
\begin{center}
\includegraphics[width=0.5\textwidth]{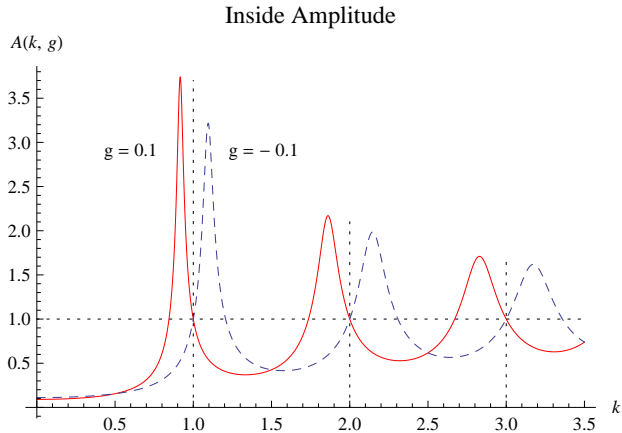}
\footnotesize
\caption{
\label{figura1}
\it Amplitude $A(k,g)$ of the eigenfunction $\psi(x;k,g)$ inside the cavity. 
Repulsive case ($g=0.1$): continuous (red) line; Attractive case 
($g=-0.1$): dashed (blue) line. Peaks become less marked with increasing order.
The dotted horizontal line represents the asymptotic value of the amplitude
at large energies.
}
\end{center}
\end{figure}

\begin{figure}[ht]
\begin{center}
\includegraphics[width=0.5\textwidth]{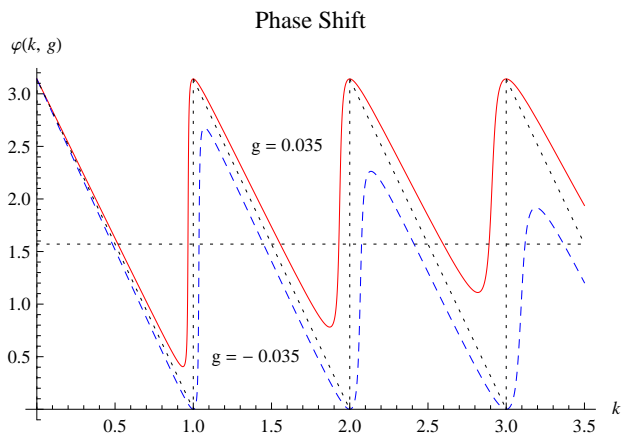}
\footnotesize
\caption{
\label{figura2}
\it Phase shift $\varphi(k,g)$ of the eigenfunction $\psi(x;k,g)$. 
Free case ($g=0$): dotted (black) line;
repulsive case ($g=0.035$): continuous (red) line; attractive case 
($g=-0.035$): dashed (blue) line. The curve for $g>0$ has been
shifted upwards by $\pi$ for comparison with the case $g<0$ at small $k$.
}
\end{center}
\end{figure}

For $g>0$ there is only a continuous spectrum, with eigenfunctions 
of the form \cite{flugge,winter,primo} (see figs.\ref{figura01} 
and \ref{figura02})
\bea
\label{diviso}
\psi(x;\,k,g) &=&
\sqrt{\frac{2}{\pi}} \frac{1}{\sqrt{4  a(k,g) b(k,g)} } 
\Big\{
\theta(\pi - x) \, \sin(k x)
\, + \,
\nonumber\\
&&  \,\,\,\,\,\,\,\,\,\,\,\,\,\,\,\,\,\,\,\,\,\,\,\,\,
+ \, \theta(x -  \pi) 
\big[
a(k,g) \, e^{ i k x} \, + \, b(k,g) \, e^{ - i k x}
\big]
\Big\} \,  
\eea
and energies
\beq
\varepsilon(k) = k^2 > 0.
\eeq
The function $\theta(y)=1$ for $y>0$ and zero otherwise is the
Heaviside step function and the square root in eq.(\ref{diviso}) 
is the arithmetical one, as $a(k,g)b(k,g)>0$ for any $k,g\in\RR$
(an over-all phase is in any case irrelevant).
The coefficients entering the eigenfunctions have the following
explicit expressions:
\bea
a(k,g) &=& - \, \frac{i}{2} \, + \, \frac{1}{4\pi g k} 
\left[ \exp( - 2 \pi i k ) - 1 \right] \, ;
\\
b(k,g) &=& + \, \frac{i}{2} \, + \, 
\frac{1}{4 \pi g k} \left[ \exp( + 2 \pi i k ) - 1 \right] \, .
\eea
These coefficients have the following two symmetries: 
\beq
\label{sim}
a(-k,g) \, = \, - b(k,g) ;
\,\,\,\,\,\,\,\,\,\,\,\,
\overline{a(k,g)} \, = \, b\left(\overline{k},\overline{g}\right) ,
\eeq
where the bar denotes complex conjugation.
The first equation says that $\psi(x;\,k,g)$ is an odd function of $k$ 
and implies that the zeroes of $a$ are opposite to those
ones of $b$, i.e. that if
\beq
b\left(k_0,g\right) \, = \, 0 ,
\eeq
then
\beq
a\left(-k_0,g\right) \, = \, 0 .
\eeq
The second equation implies that the zeroes of $a$ are the complex conjugates of 
the zeroes of $b$ for conjugate coupling, i.e. that if
\beq
b\big(k_0,g\big) \, = \, 0 ,
\eeq
then
\beq
a\left(\overline{k_0},\overline{g}\right) \, = \, 0 \, .
\eeq
For real $g$, the zeroes of $a$ are then the complex conjugates 
of the zeroes of $b$.
From the two properties above it is possible 
to reconstruct for real $g$ all the zeroes of $a$ and of $b$, 
for example, from the zeroes of $b$ in the forth quadrant.
Finally, note that for real $k$ and $g$
\beq
\overline{a(k,g)} \, = \, b\left(k,g\right),
\,\,\,\,\,\,\,\,\,\,\,\, k, \, g \in\RR,
\eeq 
so that
\beq
a(k,g) b(k,g) \, = \, |a(k,g)|^2\, = \, |b(k,g)|^2,  \,\,\,\,\,\,\,\,\,\,\,\, k, \, g \in \RR .
\eeq
In eq.(\ref{diviso}) we have assumed the standard continuum normalization:
\beq
\label{continuo}
\int_0^\infty \overline{\psi(x; \, k',g)} \, \psi(x; \, k,g) \, dx \, = \, 
\delta( k - k' ) \, ,
\eeq
with $\delta(q)$ the Dirac $\delta$-function. 
The quantity $k$ is a real quantum number but, 
since the eigenfunctions are odd 
functions of $k$, one can assume $k>0$\footnote{
The case $k=0$ has to be discarded
as one obtains in this case, because of the boundary condition,
the zero function, which is not an acceptable wavefunction.}.
Note that the spectrum is bounded from below by zero, 
uniformly in $g>0$. As we are going to show in the next section
and as anticipated in the introduction,
that is no more the case for $g<0$, because of the appearance of
a discrete spectrum.
The eigenfunctions can also be written in trigonometric form 
as \cite{flugge}:
\beq
\psi(x;k,g) \, = \, \sqrt{ \frac{2}{\pi} } \, 
\Big\{
\theta(\pi-x) \, A(k,g) \sin\big[ k \, x \big]
+
\theta(x-\pi) \, \sin \big[ k \, x + \varphi(k,g) \big]
\Big\} .
\eeq
The inside amplitude is given by (see fig.(\ref{figura1}))
\beq
A(k,g) = \frac{1}{2 \sqrt{ a(k,g) b(k,g) }  } = 
\frac{1}{\sqrt{  1 \, + \, 1/(\pi g k) \sin 2 k \pi \, 
+ \, 1/(2 \pi^2 g^2 k^2) (1 - \cos 2 k \pi) } } .
\eeq
Note that\footnote{
It may be argued that $A(k,g)$, as a function of $k$, 
resembles the shape of the total cross section of 
electron-positron annihilation into hadrons as a 
function of the c.o.m. energy $\sqrt{s}$, 
$\sigma(e^+e^-\to h's)$, above
a heavy quark-antiquark pair threshold.
Roughly speaking, the quarkonium states correspond
to the resonances of the particle inside the cavity. }
\beq
\lim_{k\to\infty} A(k,g) = 1 ,
\eeq
as expected on physical ground: high-energy states do not see 
the barrier.
The phase shift between the outside amplitude and the inside
one reads (see fig.\ref{figura2}):
\beq
\label{phase}
\cot \big[ \varphi(k,g) \big] 
\, = \, 
- \frac{\pi g k}{\sin^2(\pi k)} - \cot(\pi k) .
\eeq
For the cotangent to be invertible, let us assume
for its argument the following range:
\beq
\label{wantinvertible}
- \pi < \varphi(k,g) < 0 \,\,\,\,\,\,\,\, \mathrm{for} \,\,\,\,\, g > 0 ;
\,\,\,\,\,\,\,\,\,\,\,\,\,\,\,\,\,\,\,\,
0 < \varphi(k,g) < \pi   \,\,\,\,\,\,\,\, \mathrm{for} \,\,\,\,\,  g < 0 .
\eeq
With the above choice of the argument,
\beq
\lim_{k\to+\infty} \varphi(k,g) = 0 ,
\eeq
i.e. there is no phase shift in the high-energy limit,
in agreement with physical intuition.

\subsection{Strong-Coupling Regime}

Even though we are mostly interested to the weak-coupling
regime $|g|\ll 1$ --- to be more specific, to subtle weak-coupling
properties --- let us briefly consider the strong coupling
regime
\beq
|g| \gg 1.
\eeq
Waves find in this case a small barrier at the point $x=\pi$;
in the strong coupling limit $g\to\pm\infty$, the Dirac potential 
completely disappears and the Winter Hamiltonian becomes
then the Hamiltonian of a free particle, 
\beq
\label{eqmod0}
\hat{H}_0 = -\frac{\partial^2}{\partial x^2} ,
\eeq
on the positive axis, $x\ge 0$, with vanishing boundary conditions 
at the origin. Note also that
\beq
A(k,g) \to 1
\,\,\,\,\, \mathrm{and} \,\,\,\,\,
\varphi(k,g) \to 0
\,\,\,\,\,\,\,\,\,\,\,\,\,\, {\rm for } \,\,\,\,\,\,\, g\to\pm\infty.
\eeq
A duality therefore exists between the strong-coupling regime
of the Winter model $|g|\gg 1$ and the weak-coupling regime of a 
particle on the positive axis subjected to a small potential.
Let us remark that there is continuity in going from the
strong coupling regime, $|g|\gg 1$, to the weak-coupling one,
$|g|\ll 1$. As we are going to show in the next section,
the only singular limit is $g\to 0$.

\subsection{Weak-Coupling Regime and  Free Limit}

We are interested to the weak-coupling regime
$0< g \ll 1$ and to the free limit of the model
$g\to 0^+$ \cite{primo}.
It is clear that, unlike the previous case, 
taking this limit directly in the Hamiltonian,
which has a pole in $g=0$, is meaningless\footnote{
This is to be compared with quantum field theory,
in which the coupling to be sent to zero usually
appears in the numerator.
The $QED$ Lagrangian density, for example, reads
\beq
\mathcal{L} = - \frac{1}{4} F_{\mu\nu} F^{\mu\nu} + 
\bar{\psi} \left( i \gamma_\mu \partial^\mu - m \right) \psi 
+ e \bar{\psi} \gamma_\mu \psi A^\mu,
\eeq
where $F_{\mu\nu}$ is the field tensor and $A^\mu$ the gauge 
potential of the electromagnetic field, $\psi$ is the Dirac 
field, $e$ the electron charge, and the free limit is $e\to 0$.
}. 
We therefore take
the free limit on the eigenfunctions (which is instead meaningful) 
and then find which
Hamiltonian has the limiting eigenfunctions.

For $k$ not close to an integer within $\mathcal{O}(g)$, 
namely
\beq
\label{generic}
k \ne n + \mathcal{O}(g),
\eeq
with $n$ a non-zero integer,
\beq
|a(k,g)| \, = \, |b(k,g)| \, = \, \mathcal{O}\left( \frac{1}{g}  \right) ,
\eeq
implying that the eigenfunctions have a small amplitude
$\mathcal{O}\left(g\right) \ll 1$ inside the cavity
--- outside the cavity the amplitude is always $\mathcal{O}(1)$,
no matter which values are chosen for $k$ and $g$, because
of continuum normalization.
With the generic values of $k$ given by eq.(\ref{generic}),
outside waves are not able to excite appreciably the cavity.
In the limit $g\to 0^+$ the eigenfunctions exactly vanish
inside the cavity.

Let us then consider the phase behavior of the eigenfunctions.
Unlike the discussion on the amplitude, let us first consider
the free case.
For $g=0$, one obtains:
\beq
\varphi = \varphi_s(k,\, g = 0) = - \pi k + s \pi ,
\eeq
where $s$ is any integer.
The eigenfunctions therefore read:
\beq
\psi(x;\, k,\, g=0) \, = \, (-1)^s 
\theta(x-\pi)  \sqrt{ \frac{2}{\pi} } \sin\big[k (x-\pi)\big] , 
\, \, \, \, \, \, \, \, \, \, \, \, \, \, 
k > 0, \, \, \, \, \, k \, \ne \, n + \mathcal{O}(g) ,
\eeq
with $n$ a positive integer.
In order to satisfy eq.(\ref{wantinvertible}), 
the integer $s$ has to be a function of $k$ 
but, since an overall phase is not observable, one can take 
once and for all for example $s=0$, obtaining the usual eigenfunctions 
of a particle confined to the half-line $x\ge\pi$, with vanishing boundary
conditions, 
\beq
\psi(x=\pi;\, k, \, g=0) \, = \, 0. 
\eeq
The phase $\varphi(k,g)$ does not provide any measurable information
in the free case: it does not transfer any information
from inside the cavity to outside it.
In other words, the cavity is impermeable also as far as the
phase $\varphi$ is concerned.
Let us now consider the interacting case, $0 < g \ll 1$.
For the non-exceptional values of $k$ in eq.(\ref{generic}),
it holds
\beq
\sin(\pi k) = \mathcal{O}(1)
\eeq
and the additional term in $\varphi(k,g)$ due to the interaction
is just a small correction with respect to the free case:
\beq
\label{doublepole}
\frac{\pi g k}{\sin^2(\pi k)} \, = \, \mathcal{O}(g) .
\eeq
Therefore there is not any qualitative change of the
profile of $\varphi(k;g)$ in the region of generic $k$'s
given by eq.(\ref{generic}) (see fig.\ref{figura2}).

Let us now see what happens for $0 < g \ll 1$ and for $g\to 0^+$ 
to the amplitudes and phases of the eigenfunctions for the 
complementary values of $k$,  i.e. for the ``exceptional'' values 
\beq
\label{exceptional}
k \, \simeq \, n - n g ,
\eeq
with $n$ any non-zero integer. Since
\beq
\label{stimeaeb}
\left|a\big[n(1-g),g\big]\right| \, = \, 
\left|b\big[n(1-g),g\big]\right| \, = \, 
\frac{ \sqrt{ 1 + \pi^2 n^2 } }{2} \, |g|
\, + \, \mathcal{O}\left(g^2\right) ,
\eeq
the amplitude of $\psi(x;k,g)$ inside the cavity shows marked
peaks. In the limit $g \to 0^+$
\beq
\left|a\big[n(1-g),g\big]\right| \, = \, \left|b\big[n(1-g),g\big]\right| 
\, \to \, 0 
\eeq
and the inside amplitude diverges.
This divergence, which (at face value) is physically meaningless, actually signals
a qualitative change of the spectrum. 
To have a finite amplitude inside the cavity for $g\to 0$, one has to impose  
that the eigenfunctions exactly vanish outside it, necessarily obtaining 
states with a finite normalization. Therefore a discrete spectrum emerges 
out of the continuum one for the exceptional $k$-values in 
eq.(\ref{exceptional}) when
\beq
g \to 0 \, \, \, \, \, \,\,\,\,\,\,\,
{\rm so} \,\,\, {\rm that} \,\,\, \, \, \,\,\,\, k \simeq n(1-g) \to n .
\eeq
By normalizing these eigenfunctions to one, we obtain:
\beq
\sqrt{ \frac{2}{\pi} } \, \theta(\pi-x) \, \sin( n x ) ,
\eeq
with $n$ a positive integer. The above ones are the eigenfunctions
of a particle confined to the segment $0\le x \le \pi$.

As far as the phase behavior is concerned, the term on the l.h.s. of 
eq.(\ref{doublepole}) has a double pole for $k \to \, \mathrm{integer}$,
while the $g$-independent term in eq.(\ref{phase}),  $\cot(\pi k)$,
only has a simple pole.
The consequence is that the term proportional to $g$ is not uniformly small
in $k$ even for $|g|\ll 1$ and actually dominates for $k\to$ integer.
In the region specified by eq.(\ref{exceptional})
\beq
\varphi \, \simeq \, \mp \frac{\pi}{2}
\eeq
for $g>0$ and $g<0$ respectively.
Furthermore, $\varphi(k,g)$ passes through $\mp\pi/2$ with a high slope
when $k$ passes through $n(1-g)$:
\beq
\label{Dfase}
\left.\frac{\partial\varphi}{\partial k}\right|_{k=n(1-g)} =  
\frac{\pi}{1+\pi^2 n^2} \, \frac{1}{g^2}
+  \mathcal{O}\left(\frac{1}{g}\right) .
\eeq
This sudden phase variation is a typical resonant behavior
\cite{scatteringth}.
An important point is that, unlike the free case, $\varphi(k; g \ne 0)$ 
is a continuous function of $k$ and is a measurable.
In the limit $g\to 0$ the phase $\varphi(k,g)$ develops infinite slopes 
at the integers $k=n$ (see fig.\ref{figura2}).

By looking at eqs.(\ref{exceptional}), (\ref{stimeaeb}) and (\ref{Dfase}), 
one may observe that the ``effective coupling'' of the $n$-th
resonance to the continuum rather than $g$ is actually 
$\approx n g$. 
That implies that one has to face strong-coupling phenomena for
large $n$ even when $|g|\ll 1$, because the perturbative 
expansion involves powers of $n g$ rather than powers of $g$:
we will make these considerations more precise in later
sections.

We can summarize the above findings by saying that, in the free limit,
the system described by the Winter Hamiltonian decomposes into two
non-interacting subsystems.
The first subsystem is particle in a box, with Hamiltonian
\beq
\hat{H}_1 = -\frac{\partial^2}{\partial x_1^2}, 
\, \, \, \, \, \, \, \, \, \, \,\,\,\,\,
0 \le x_1 \le \pi,
\eeq
with vanishing boundary conditions:
\beq
\psi_1(x_1=0,\,t) = \psi_1(x_1=\pi,\,t) = 0 , \,\,\,\, \, \, \, \,\,\,\,
t \in \RR.
\eeq
As well known, this system has an infinite tower of
discrete states,
\beq
\psi_1^{(n)}(x_1) =  \sqrt{ \frac{2}{\pi} } \sin \left( n x_1 \right) ,
\eeq
with $n\in\NN_+$, and no continuum spectrum.
The second subsystem is a particle in a half-line,
with Hamiltonian
\beq
\hat{H}_2 = -\frac{\partial^2}{\partial x_2^2},
\, \, \, \, \, \, \, \, \, \,\,\, 
\pi \le x_2 < \infty,
\eeq
with wavefunctions vanishing at the only boundary point $x=\pi$:
\beq
\psi_2(x_2=\pi,\,t)=0 , \, \, \, \, \, \, \,\,\, t \in \RR .
\eeq
$\hat{H}_2$ has a continuous spectrum only, with eigenfunctions
(normalized to a $\delta$-function) of the form
\beq
\psi_2\left(x_2;k \right) 
= \sqrt{ \frac{2}{\pi} } \sin\left[k (x_2-\pi)\right] ,
\eeq
with $k>0$.
Therefore Winter Hamiltonian $\hat{H}(g)$ factorizes in the limit
$g\to 0^+$ in the sum of the Hamiltonian of a particle in a box
and the Hamiltonian of particle in a half-line:
\beq
\lim_{g\to 0^+ }\hat{H}(g) = \hat{H}_1 + \hat{H}_2 .
\eeq

\subsection{Resonances}

To study the decay, as well as the formation, of metastable states,
it is convenient to introduce generalized
eigenfunctions $\psi(x;\,k,g)$ with $k\in\CC$, called resonances or antiresonances, 
which satisfy purely outgoing or purely incoming boundary conditions, 
i.e.
\beq
\label{BC}
\left( \frac{d}{dx} - i k \right) \psi(x;\,k,g) = 0
\,\,\,\,\,\,\,\,\,\, \mathrm{for} \,\, \,\,\, x>\pi, 
\eeq
with
\beq
\mathrm{Re} \, k > 0 
\,\,\,\,\, \mathrm{or} \,\,\,\,\,  
\mathrm{Re} \, k < 0
\eeq
respectively.
Let us remark that while in the case of the Winter model the boundary condition above, implying
\beq
\psi(x;\,k,g) \, \approx \, e^{i k x} , 
\eeq
can be imposed at any point outside the cavity, for a general short-range 
potential, one has to impose it at $x\to+\infty$.
By equating to zero the coefficient $b(k,g)$ of the component $\exp(-ikx)$
in $\psi(x;\,k,g)$ (see eq.(\ref{diviso})), 
we obtain the transcendental equation in $k$
\beq
\label{bzero}
\exp( 2 \pi i k ) 
\, + \, g 2 \pi i k \, - 1 \, = \, 0 ,
\eeq
having a countable set of solutions
$\{k^{(n)}(g)\}$ lying in the lower half of the $k$-plane, 
\beq
\mathrm{Im} \, k^{(n)}(g)  \, < \, 0
\eeq
and reading for $|g|\ll 1$:
\beq
\label{insert1}
k^{(n)}(g) \, = \, n ( 1 - g + g^2 ) - i \pi  n^2 g^2
+ \mathcal{O}\left( g^3  \right) ,
\eeq
where $n$ is any non-zero integer.
For $n>0$, the zeroes lie in the forth quadrant and are associated
to resonances, while for $n<0$ they lie in the third quadrant
and are associated to antiresonances.
In general, the functions $k^{(n)}(g)$'s with $n\ne 0$ are 
defined as the zeroes of the transcendental equation above, 
\beq
b\left[k^{(n)}(g), g \right] \equiv 0 ,
\eeq
satisfying the initial condition
\beq
k^{(n)}(g=0) \, = \, n .
\eeq
These functions are fundamental elements of the Winter model.
In the next section we will present a higher-order perturbative expansion
for $k^{(n)}(g)$ --- as already noted, we will find that the expansion 
parameter, rather than $g$, is actually $g n$.
In sec.(\ref{sect5}), in order to fully understand their properties, 
we will analytically continue the $k^{(n)}(g)$'s for complex $g$.  
All the $k^{(n)}(g)$'s will turn out to be the various branches 
of the multivalued function $k = k(g)$ implicitly defined
by $b\left[k(g), g \right] \equiv 0$.

Because of the first symmetry between the functions $a(k;g)$
and $b(k;g)$ discussed in the previous section, the zeroes
$\{\chi^{(n)}(g)\}$ of the function $a(k;g)$ 
have a positive imaginary part for any $n\ne 0$. 
They can be defined as
\beq
\chi^{(n)}(g) \, \equiv \, - k^{(n)}(g) \, = \, 
- n ( 1 - g + g^2 ) + i \pi  n^2 g^2 + \cdots
\eeq
and are associated to resonances for $n>0$ (second quadrant) 
and to antiresonances for $n<0$ (first quadrant).
Roughly speaking, because of the first symmetry between the coefficients,
all the dynamical information is already contained in just one coefficient;
we have therefore considered explicitly only the coefficient $b(k;\,g)$.
By using also the second symmetry, one can easily show that for $g\in\RR$ 
the real part of $k^{(n)}(g)$ is odd in $n$, while the imaginary part
is even in $n$. From the relation
\beq
\chi^{(n)}(g) \, = \, \overline{ k^{(-n)}(g) }
\eeq
it follows indeed
\beq
k_1^{(n)}(g) + i  k_2^{(n)}(g) \, = \, - k_1^{(-n)}(g) + i  k_2^{(-n)}(g),
\eeq
where $k^{(n)}(g) = k_1^{(n)}(g) + i  k_2^{(n)}(g)$ with  
$k_1^{(n)}(g), \, k_2^{(n)}(g)\in\RR$. 
For real $g$ the Winter Hamiltonian is real
and therefore time-reversal invariant, so that the formation
and the decay of a resonance occur in the same way. 

According to the formula above, for $n=0$ one would obtain $k^{(0)}(g) \equiv 0$;
unlike the other cases, the function $k^{(0)}(g)$ does not posses a convergent 
expansion in powers of $g$.
Relevant expansions in this case are an expansion in powers
of $g+1$ for $|g+1| \ll 1$
\beq
\label{gplusone}
k^{(0)}(g) \, \simeq \, 
\frac{i}{\pi}(g+1) 
\, + \, \frac{2 i}{3\pi} (g+1)^2 
\, + \,\mathcal{O}\left[ (g+1)^3 \right] \, , 
\eeq
as well as an expansion involving exponentials and poles for $g<0$, $|g| \ll 1$:
\beq
\label{gexps}
k^{(0)}(g) \, = \, - \, \frac{i}{2\pi g} 
\Big[ 1 - e^{1/g} + \mathcal{O}\left( e^{2/g}  \right) \Big]  \, .
\eeq
We will see the physical significance of these expansions in
the next section.
Let us remark that $k^{(0)}(g)$ is a purely imaginary number 
for $g\in\RR^-$, with a positive imaginary part for
\beq
- 1 < g < 0 .
\eeq
The latter case corresponds to an exponentially decaying wavefunction
for $x\to+\infty$ with negative energy, i.e. a bound state.

\noindent
Coming back to the resonances, we obtain for their wavefunctions
($n> 0$): 
\bea
\label{thetafun} 
\theta^{(n)}(x,t;g) &\equiv& \sqrt{ \frac{2}{\pi} } 
\left\{
\theta(\pi-x) \, \sin \left[ k^{(n)}(g) x \right] 
+ \theta(x-\pi) \, 
\frac{ \pi g  k^{(n)}(g) }{ 2 \pi i g  k^{(n)}(g) - 1 }
\, e^{ i \, k^{(n)}(g) \, x }
\right\} \times
\nonumber\\
&& \,\,\,\,\,\,\,\,\,\,\,\,\,\,\,\,\,\,\,\,\,\,\,\,\,\,\,\,
\,\,\,\,\,\,\,\,\,\,\,\,\,\,\,\,\,\,\,\,\,\,\,\,\,\,\,\,\,
\,\,\,\,\,\,\,\,\,\,\,\,\,\,\,\,\,\,\,\,\,\,\,\,\,\,\,
\,\,\,\,\,\,\,\,\,\,\,\,\,\,\,\,\,\,\,\,\,\,\,\,\,\,\,
\times \exp\left[ - i \, \varepsilon^{(n)}(g) \, t \right]
\eea
and for the (complex) energies
\beq
\varepsilon^{(n)}(g) \, = \, k^{(n)}(g)^2 .
\eeq
The resonances evolve diagonally in time by means of the factors 
\beq 
\label{timefact}
E^{(n)}(t;\,g) \, \equiv \, 
\exp\left[ - i \, \varepsilon^{(n)}(g) \, t \right] \, = \, 
\exp\left[ - i \, \omega^{(n)}(g) \, t - \, \frac{1}{2} \Gamma^{(n)}(g) t \right] \, .
\eeq
Since the energies are complex for $g\ne 0$, on the last member 
we have split them into real and imaginary parts as:
\beq
\varepsilon^{(n)}(g) \, = \, \omega^{(n)}(g) \, - \, \frac{i}{2} \Gamma^{(n)}(g) ,
\eeq
where $\omega^{(n)}(g)$ is the frequency and $\Gamma^{(n)}(g)$ is the decay
width of the resonance $n$:
\bea
\omega^{(n)}(g) &=& {\rm Re } \left[ k^{(n)}(g)^2 \right] \, = \, n^2(1-2g) 
\, + \, \mathcal{O}\left(g^2\right);
\\
\Gamma^{(n)}(g) &=& - 2 \, {\rm Im }\left[ k^{(n)}(g)^2 \right] \, = \, 4\pi g^2 n^3 
\, + \, \mathcal{O}\left(g^3\right).
\eea
Note that $\omega^{(n)}(g)$ is even in $n$, while $\Gamma^{(n)}(g)$
is odd in $n$, and that $E^{(n)}(t=0;\,g)=1$. 
Let us make a few remarks. 
\begin{enumerate}
\item
The exponential decay of a resonance with time is not in contradiction with 
the conservation of probability (i.e. of the number of particles) 
because $\theta^{(n)}(x, t ; g)$, as a function of $x\in\RR^+$, is not 
a normalizable wavefunction and describes an outgoing 
flux of particles at $x\to+\infty$
\footnote{
For real $g$, the Winter Hamiltonian is hermitian only with boundary 
conditions implying no net flux of particles at infinity.
}.
Actually, a resonance wavefunction is not even a bounded function --- like
the ordinary eigenfunctions--- and diverges 
exponentially for $x\to+\infty$,\footnote{
In the large-time numerical evolution of wavepackets initially concentrated
between potential wells, such an exponential increase is actually observed 
in a large space interval \cite{refbase,refbase2}.
} 
since $\mathrm{Im} \, k^{(n)}(g) < 0$  \cite{gamow}.
By writing indeed
\beq
k \, = \, k_1 - i k_2, \,\,\,\,\,\,\,\, k_1, \, k_2 \in \RR ,
\eeq
the outgoing boundary condition and the positivity of the width, 
\beq
k_1 \, > \, 0, \,\,\,\,\,\,\,\, 
\Gamma \, = \, 4 k_1 k_2 > 0 ,
\eeq
imply $k_2>0$, so that the resonance wavefunction diverges exponentially
for $x\to+\infty$ as
\beq
e^{i k x} \, = \,  e^{i k_1 x + k_2 x} ;
\eeq
\item
The resonances, just like the ordinary eigenfunctions, are smooth functions
in $\RR^+\backslash\{\pi\}$, while they are only continuous at $x=\pi$.
Even a finite discontinuity would indeed produce an infinite average
kinetic energy (see next section);
\item
Outside the cavity, $\theta^{(n)}(x,t;g)$ is $\mathrm{O}(g)$ compared to
the inside, because of the explicit $g$ factor in the second term on the 
r.h.s. of eq.(\ref{thetafun} ). As expected on physical ground, apart
from the (slow for $g\ll 1$) exponential divergence for $x\to+\infty$ 
discussed above, the wavefunction is concentrated inside the cavity.
\end{enumerate}

\subsection{Attractive Case}

Even though we are mostly interested to the weakly repulsive case,
i.e. to $0 < g \ll 1$, let us briefly discuss the modifications of the
spectrum in the attractive case, i.e. for $g<0$  \cite{primo}.
In the latter case, there is a continuous spectrum as in the repulsive 
case, while for
\beq
- 1 < g < 0
\eeq
there is also a discrete spectrum consisting of a single bound
state. The discrete  $(d)$ eigenfunction reads:
\beq
\psi_{d}(x;\,g) =
C_{g} 
\left[
\theta(\pi - x) \left( e^{ \, \chi(g) \, x} -  e^{- \chi(g) \, x} \right)
\,+\,\theta(x -  \pi) \left( e^{ 2 \pi \chi(g) } - 1 \right) e^{- \chi (g)\, x}
\right]
\eeq
and has the negative energy
\beq
\varepsilon_{d}(g) \, = \, - \chi^2(g) \, < \, 0 .
\eeq
The quantity $\chi(g)\in\RR^+$ is the imaginary part of the bound-state solution 
of the equation $b(k,g)=0$, which is purely imaginary, 
\beq
k^{(0)}(g) \, = \, i \chi(g) .
\eeq
It is the positive solution of the transcendental equation
\beq
\label{trascend}
e^{- 2 \pi \chi(g) } \, = \, 1 +  2 \pi g \chi(g) \, .
\eeq
By normalizing the wavefunction to one, the constant above reads:
\beq
C_{g} \, = \, 
\sqrt{ \, \frac{ \chi(g) }{ e^{2 \pi \chi(g)  } - 1 - 2 \pi \chi(g) } ~ } \, .
\eeq
For a ``loosely-bounded'' particle, i.e. for $|1+g| \ll 1$, 
the transcendental equation above has the approximate solution 
(see eq.(\ref{gplusone}))
\beq
\chi(g) \, \simeq \, 
\frac{1}{\pi}(g+1) \, + \, 
\frac{2}{3\pi} (g+1)^2 \, + \, 
\mathcal{O}\left[ (g+1)^3 \right] \, , 
\eeq
while for a ``tightly-bounded'' particle, i.e. for a negative coupling 
of small size, $g<0$, $|g| \ll 1$, one has the expansion 
(see eq.(\ref{gexps}))

\beq
\chi(g) \, = \, - \, 
\frac{1}{2\pi g} 
\Big[ 1 - e^{1/g} + \mathcal{O}\left( e^{2/g}  \right) \Big]  \, .
\eeq
We will rederive the above expansions when studying the
analytic continuation of the functions $k^{(n)}(g)$ controlling the
resonance behavior of Winter model (see later).
In the case $|g|\ll 1$ (tight binding), by omitting exponentially 
small terms, we have the explicit formula:
\beq
\psi_{d}(x;\,g) \, \simeq \, 
\frac{1}{\sqrt{2\pi |g|}}
\left[
\theta(\pi - x)  
\exp\left( \frac{x-\pi}{ 2 \pi |g| } \right)
\, + \, \theta(x -  \pi) 
\exp\left( \frac{\pi-x}{ 2 \pi |g| } \right)
\right]
\eeq
and
\beq
\varepsilon_{d}(g) \simeq  - \frac{1}{ 4 \pi^2 g^2 } . 
\eeq
As discussed in the introduction, for $g\to 0^-$ the spectrum 
becomes unbounded from below, while it is uniformly bounded by 
zero for $g\to 0^+$.

Let us now consider the free limit in the (more complicated) 
attractive case, $g\to 0^-$.
Roughly speaking, in the this case, factorization 
is not so clean because of the presence of a discrete spectrum.
As we have seen, for $g\to 0$ the continuous-spectrum eigenfunctions 
leave the cavity for generic $k$ values, while they concentrate inside 
it for the exceptional $k$ values $k\, \approx \,\mathrm{integer}$.
On the contrary, the bound-state wavefunction extends symmetrically 
on both sides of the potential wall for $g\to 0^-$ (see fig.\ref{figura_bs}).
However, since for $g\to 0^-$ the wavefunction is concentrated
in a thin layer around $x=\pi$, of width $\approx 4\pi |g|$, the 
resulting coupling between the box $[0,\pi]$ and the half-line $[\pi,\infty)$ 
is small. In a weak sense indeed:
\beq
\lim_{g\to 0^-} \psi_d^2(x;g) \, = \, \delta(x-\pi) ,
\eeq
while
\beq
\lim_{g\to 0^-} \psi_d(x;g) \, = \, 0 .
\eeq

\section{Temporal Evolution of Metastable States}
\label{sect3}

We study the forward time evolution, $t\ge 0$, of wavefunctions $\psi^{(l)}(x,t;g)$ 
which coincide at the initial time, $t=0$, with the box eigenfunctions and vanish 
outside it \cite{winter,primo,secondo}:
\beq
\label{initial}
\psi^{(l)}(x,t=0;\,g) \, = \, 
\left\{
\begin{array}{cc}
\sqrt{ 2/\pi } \, \sin \left( l \, x \right) & {\rm for} ~ 0 \le x \le \pi ; ~~
\\
0 &{\rm for} ~ \pi < x < \infty \, ,
\end{array}
\right.   
\eeq
where $l$ is a positive integer.
The above wavefunction is just a wavepacket with support\footnote{
The support of a numerical function $f:D\to\CC$ is the closure of the set where
the function is not vanishing: 
$\mathrm{Supp} f \equiv \overline{\{x\in D : f(x)\ne 0 \}}$. }
entirely contained inside the cavity.
Physically, that corresponds to consider an initial state 
containing an unstable particle such as, for example, a $Z^0$, but not any 
of its decay products.
By means of a spectral representation in (ordinary) eigenfunctions and 
contour deformation in the $k$-plane, the wave-function at time $t>0$ 
can be exactly written as \cite{primo}:\footnote{
For $-1<g<0$ there is also an additional term 
on the r.h.s. of eq.(\ref{defVtheta}) coming from the
discrete spectrum, a case which we do not consider explicitly.}
\beq
\label{defVtheta}
\psi^{(l)}(x,t;g) =  
\sum_{n=1}^{\infty} V(g)_{l n} \, \theta^{(n)}(x,t;g) \, + \, P^{(l)}(x,t;\,g)\, ,
\eeq
where $\theta^{(n)}(x,t;g)$ is the $n$-th resonance wavefunction constructed
in the previous section and $P^{(l)}(x,t;\,g)$ is a non-exponential (power-like)
contribution, having the exact integral representation 
\bea
P^{(l)}(x,t;\,g) &\equiv& 
e^{-i\pi/4} \left(\frac{2}{\pi}\right)^{3/2} 
\left\{ \theta(\pi - x) 
\int\limits_{0}^{\infty} q^{(l)}\left( k \, e^{-i\pi/4}; \, g \right) 
 \sin\left( k \, e^{-i\pi/4} x \right) \, e^{-k^2 t} \, dk \, + \right. 
\nonumber\\
&&\,\,\,  + \, \theta(x-\pi) 
\left[
\int\limits_{0}^{\infty} q^{(l)}\left( k \, e^{-i\pi/4};\, g \right) 
a(k\, e^{-i\pi/4} ; \, g) \, e^{ \exp(i\pi/4) k x - k^2 t} \, dk \, + \right.
\nonumber\\
&&\,\,\,\,\,\,\,\,\,\,\,\,\,\,\,\,\,\,\,\,\,\,\,\, + \, \left. \left. 
\int\limits_{0}^{\infty} q^{(l)}\left( k \, e^{-i\pi/4};\, g \right) 
b(k\, e^{-i\pi/4} ; \, g) \, e^{ - \exp(i \pi/4) k x - k^2 t} \, dk 
\right]
\right\},
\eea
with
\bea
q^{(l)}(k;\,g) &\equiv&  \frac{(-1)^l l \sin k \pi}{ k^2 - l^2 } 
\frac{1}{ 4 a(k;\,g) \, b(k;\,g) }
\nonumber\\
&=&  \frac{(-1)^l l \sin k \pi}{ k^2 - l^2 } 
\, \frac{1}{ 1 + 1/(\pi g k) \sin 2 k \pi + 1/(2 \pi^2 g^2 k^2) ( 1 - \cos 2 k \pi ) } .
\eea
The dependence on the initial state, i.e. on $l$,
is contained in the first factor on the second and last member of
the above equation, while the dependence on the dynamics is contained
in the second factor.
The following asymptotic expansion for $t \gg 1$ holds:
\beq
P^{(l)}(x,t;\,g) \approx
\frac{ e^{i\pi/4} }{\sqrt{2}} \frac{ (-1)^l }{l} \frac{g}{1+g} 
\left[
\theta(\pi-x) \frac{g \, x}{1+g} + 
\theta(x-\pi) \left( x -\frac{\pi}{1+g} \right)
\right] \frac{1}{ t^{3/2} }
+ \mathcal{O} \left( \frac{1}{t^{5/2} } \right).
\eeq
The following remarks are in order:
1) the above function is continuous in $x=\pi$, with a discontinuous first derivative
at the same point; 
2) the power corrections are $\mathcal{O}\left(g^2\right)$ inside the cavity
and $\mathcal{O}\left(g\right)$ outside it, implying larger contributions
in the latter case for $|g|\ll 1$. 
Power contributions however vanish in both cases for $g\to 0$, 
while for $g\to\pm\infty$ they represent typical dispersive behavior.
As we are going to show in the non-perturbative section, in the strong-coupling 
limit the resonance contributions indeed disappear, because the functions 
$k^{(n)}(g)$ have imaginary parts $\to - \, \infty$ \cite{primo}.

In the following we shall not concentrate on the post-exponential power-corrections 
in time related to $P^{(l)}(x,t;\,g)$, which do not have a resonance 
interpretation and therefore are not relevant for the present discussion.
Physically, they involve the emission of very low-energy particles out of 
the cavity at very large times.
We will also take $t$ sufficiently large to avoid the considerations
of the pre-exponential effects.
The latter are related to a fast rearrangement of the initial wavefunction, 
which modifies its short-wavelength components in order to ``fit'' inside the 
cavity \cite{winter}. For $0< g \ll 1$ there is a large temporal window 
where the exponential decay is a good approximation inside the cavity 
\cite{primo,secondo}: 
\beq
\label{exptime}
1 \, \ll \, t \, \lsim \, \frac{\log(1/g)}{g^2} .
\eeq

Finally, the quantity $V(g)$ is an infinite matrix describing the 
coupling of the initial state to the resonances, with entries 
\beq
\label{defV1}
 V(g)_{l n} \, \equiv  \,       
\frac{ (-1)^{l+1} \, 2 l  } { k^{(n)}(g)^2 - l^2 } \, 
\frac{ g \, k^{(n)}(g) \, \exp\left[ i \pi k^{(n)}(g) \right] }
{ \exp\left[ 2 \pi i k^{(n)}(g) \right] + g } \, ,
\eeq
where in the first factor on the r.h.s. we have isolated the
dependence on the initial state, i.e. on $l$.
As we are going to explicitly show in the next sections, 
the matrix elements $V(g)_{l n}$, even for very small $g$, decay quite 
slowly for $l,n\to\infty$.
However, at fixed $x$ and $t>0$, that does not produce any convergence 
problem on the r.h.s. of eq.(\ref{defVtheta}), since the resonances 
$\theta^{(n)}(x,t;g)$'s decay exponentially with $n$.
Their widths, $\Gamma^{(n)}(g)$'s, grow indeed faster with $n$ than the 
imaginary part of the $k^{(n)}(g)$'s 
(see eqs.(\ref{insert3}) and (\ref{larghezzeNP})):
\beq
\Gamma^{(n)}(g) \, \approx \, n \log n
\,\,\,\,\,\,\,\,\, \mathrm{while} \,\,\,\,\,\,\,\,\,\,\,
\mathrm{Im} \, k^{(n)}(g) \, \approx \, \log n
\,\,\,\,\,\,\,\,\,\,\,\,\,\,\,\, (n\gg 1 ). 
\eeq
In other words, the series on the r.h.s. of eq.(\ref{defVtheta})
is convergent even for a slow decay of $V(g)_{l n}$ for $n\to\infty$
as long as $t>0$.

\subsection{Resonance Mixing}

Let us now discuss in general terms the content of eq.(\ref{defV1}), 
which is the second fundamental element of the Winter model.
In the next sections we will be more concrete by inserting 
explicit approximations, holding in different regimes, for the 
$k^{(n)}(g)$'s inside of $V(g)$.
Eq.(\ref{defV1}) is indeed an exact relation for the matrix entries
of $V(g)$ in terms of the coupling $g$ and of the functions 
$k^{(n)}(g)$'s.
Unlike the dependence on $n$, which involves the detailed form of $k^{(n)}(g)$,
the dependence on $l$, i.e. on the initial state, of $V(g)_{l,n}$ is quite explicit; 
in particular:
\beq
V(g)_{l,n} \approx \frac{1}{l}
\,\,\,\,\,\,\,\,\, {\rm for } \,\,\,\,\, l \gg n . 
\eeq
Since the functions $k^{(n)}(g)$ are continuous in $g=0$ (being analytic),
the diagonal elements $V(g)_{n,n}$ are enhanced for $|g|\ll 1$, because
in this case $k^{(n)}(g)\approx n$ and a small denominator arises.
That is in agreement with physical intuition: one expects 
the $l$-th initial wavefunction in eq.(\ref{initial}) to excite
mostly the $l$-th resonance.
A fundamental point is however that the above property is not exact:
the initial box eigenfunctions do not excite a single resonance at a time, 
but in principle the whole spectrum, with strengths given by the elements 
of the matrix $V(g)$.
Since the $k^{(n)}(g)$'s, being analytic functions,
cannot have non-isolated zeroes, it follows that the $V(g)_{l,n}$'s
are all non-zero for most of the values of $g$.
Realistic physical models have many excitations\footnote{
Just think to the QCD description of the collisions at the LHC, where 
dozens of hadrons are created at each event!
}
so, in order to have real control on the system, one is faced with the
problem of constructing initial states which excite, as much as possible, 
a single state at a time.
In experimental high-energy physics, for example, that is the problem 
of producing monochromatic beams of specified particles,
such as antiprotons or neutrinos, out of collisions.
In lattice QCD, that involves the construction of a good
interpolating field, exciting out of the vacuum, within human 
capabilities, only the states one is interested to.
In essence, the problem is that of eliminating, as far as
we can, the backgrounds.
Let us stress that without some background subtraction
there is no point even in building up experiments.
In our case, to excite one resonance at a time, an idea could be to 
construct a new initial wavefunction, i.e. a new wavepacket, by inverting 
the infinite matrix $V(g)$ and applying it to the box eigenfunctions.
We then go from
\beq
\psi^{(l)}(x,t=0;\,g) \, \equiv \, \sqrt{\frac{2}{\pi}} \, \theta(\pi-x) \, \sin(l x) 
\eeq
to
\beq
\label{rotatebis}
\Phi^{(l)}(x,t=0;\,g) \, \equiv \, \sqrt{\frac{2}{\pi}} \, \theta(\pi-x) 
\, \sum_{n=1}^\infty \big(V(g)^{-1}\big)_{l n} \sin(n x) .
\eeq
That way one obtains a formal initial wavefunction evolving, as far as
exponential evolution is concerned, exactly as a single resonance:
\beq
\label{ancora}
\Phi^{(l)}(x, t ; g) \, = \, \theta^{(l)}(x,t;g) 
+ \Pi^{(l)}(x, t ; g), 
\,\,\,\,\,\,\,\,\,\, t > 0,
\eeq
where
\beq
\Pi^{(l)}(x, t ; g) \, \equiv \, 
\sum_{n=1}^\infty \big(V(g)^{-1}\big)_{l n} \, P^{(n)}(x,t;\,g).
\eeq
The ``experimental meaning'' of eq.(\ref{rotatebis}) is rather transparent:
in order to excite the $l$-th resonance only and then observe a "diagonal" 
time evolution as in the free case ($g=0$), one should prepare the initial 
state as the coherent superposition of free eigenfunctions given at its r.h.s..
To define invertibility for an infinite matrix (and
eventually give a rule for explicitly computing the inverse) requires in 
general a delicate limiting procedure. We will deal with this issue in the 
next section by means of perturbation theory and we will give some 
results concerning the exact inverse in the last section of the paper.  
The main problems in the above strategy to identify the excitations
of the Winter model, i.e. in eq.(\ref{rotatebis}), are however the following.
As already discussed, $\theta^{(l)}(x,t;g)$ diverges exponentially 
for $x\to+\infty$, so it is not even a bounded function,
\beq
\theta^{(l)}(x,t;g) \, \notin \, L^\infty\left(\RR^+\right) ,
\eeq
while the ``counter-rotated'' initial wavefunction $\Phi^{(l)}(x,t=0;\,g)$, 
being a wavepacket with support inside the cavity, is supposed to be 
square summable,
\beq
\Phi^{(l)}(x,t=0;g) \, \in \, L^2\left(\RR^+\right).
\eeq
However exact time evolution preserves the norm of the states, so
$\Phi^{(l)}(x,t;g)$ should be square summable at all times:\footnote{
In principle, non-exponential corrections could invalidate
this reasoning but, at the present level of understanding,
that does not seem to be the case.}
\beq
\Phi^{(l)}(x,t;g) \, \in \, L^2\left(\RR^+\right).
\eeq
On physical ground, it seems difficult to completely "fill'' a resonant 
state by evolving a properly constructed wavepacket.
The physical idea behind the use of resonant states
to describe the long-time evolution of {\it specific} wave-packets (i.e. initially 
concentrated between potential wells), is indeed to consider a limited 
region of space, even though the problem is formally defined on the whole 
positive half-line. 
On the technical side, the problem of the above strategy is that the 
matrix elements $V(g)_{l n}$, even for very small $g$, decay quite 
slowly for $l,n\to\infty$.
By inverting the matrix $V(g)$, one obtains 
an infinite matrix $V(g)^{-1}$ with elements $\left(V(g)^{-1}\right)_{l n}$ also slowly 
decaying for $l,n\to\infty$ --- at least, that is what happens in perturbation theory
and what exact (numerical) computations point out (see next sections).
When one applies the matrix $V(g)^{-1}$ to the infinite vector containing the
initial data, 
\beq
\sqrt{\frac{2}{\pi}} \, \theta(\pi-x) \big( \sin x,  \sin 2x, \sin 3x, \cdots \big) ,
\eeq
singular initial wavefunctions are obtained.
A final remark about the formal construction above is in order. 
Eq.(\ref{rotatebis}) represents the initial wavefunction
$\Phi^{(l)}(x,t=0;g)$ as a sine Fourier transform with coefficients
$(V(g)^{-1})_{l n}$:
the set $\{\sin n x, n\in\NN\}$ forms indeed a complete orthogonal basis 
on the interval $[0,\pi]$ \cite{kolmogorov}.  
It is clear that one could have taken any other basis, as the
results are independent on that (arbitrary) choice.

In view of the above facts and considerations, it is natural to take a 
more "modest'' approach to identify the excitations of Winter model,
consisting in "counter-rotating" the low-energy resonances only. 
Let us then assume that one is interested to the first $N$ resonances only and
rewrite eq.(\ref{defVtheta}) in the form
\beq
\label{defVthetafin}
\psi^{(l)}(x,t;g) =  
\sum_{n=1}^{N} W(g)_{l n} \, \theta^{(n)}(x,t;\,g) \, + \, r^{(l)}(x,t;\,g,N) 
\, + \, P^{(l)}(x,t;\,g) \, ,
\eeq
where $1 \le l \le N$ and $W(g)$ is an $N\times N$ matrix with entries 
equal to the corresponding ones in $V(g)$:
\beq
W(g)_{l n} \, \equiv \, V(g)_{l n} ,
\,\,\,\,\,\,\,\,\,\, l,n=1,2,\cdots,N,
\eeq
while
\beq
\label{funzioneresto}
r^{(l)}(x,t;\,g,N) \, \equiv \, \sum_{n=N+1}^{\infty} V(g)_{l n} \, \theta^{(n)}(x,t;g) .
\eeq
The latter is a remainder function containing the high-energy resonances
which one does not want to treat explicitly.
Since the lifetimes of the resonances quickly decay with increasing 
energy, we expect that neglecting the remainder function on the r.h.s. of 
eq.(\ref{defVthetafin}) should be a reasonable approximation for $t$ sufficiently large
--- but of course still within the exponential time region specified by eq.(\ref{exptime}).
Furthermore, since $n\ne l$ on the r.h.s. of eq.(\ref{funzioneresto}), 
$r^{(l)}(x,t;\,g,N)$ involves small 
$\mathcal{O}(g)$ off-diagonal couplings of the resonances to the $l$-th state
(see next section).
The meaning of the truncation at $N$ should be rather clear. 
For $N=1$, for example, one explicitly considers the first resonance only;
an example of this approximate scheme with $N=2$ has been sketched in \cite{secondo}.
We then consider the evolution of the wavefunction $\phi^{(l)}(x,t;\, g,N)$ 
given at the initial time, $t=0$, by:
\beq
\phi^{(l)}(x,t=0;\,g,N) \, \equiv \, \sqrt{\frac{2}{\pi}} \, \theta(\pi-x) 
\, \sum_{n=1}^N \big(W(g)^{-1}\big)_{l n} \sin(n x) .
\eeq
Its evolution reads:
\beq
\phi^{(l)}(x,t;\,g,N) \, = \, \theta^{(l)}(x,t;g) \, + \, \rho^{(l)}(x,t;\,g,N) 
\, + \, \pi^{(l)}(x,t;\,g,N) ,
\eeq
where
\bea
\rho^{(l)}(x,t;\,g,N) &\equiv& \sum_{n=1}^N \big(W(g)^{-1}\big)_{l n} r^{(n)}(x,t;\,g,N)
\nonumber\\
&=& \sum_{s=N+1}^\infty R_{l s}(g,N) \, \theta^{(s)}(x,t;g),
\eea
with
\beq
R_{l s}(g,N) \equiv \sum_{n=1}^N \big(W(g)^{-1}\big)_{l n} V(g)_{n s}  
\eeq
and
\beq
\pi^{(l)}(x,t;\,g,N) \, \equiv \, \sum_{n=1}^N \big(W(g)^{-1}\big)_{l n} \, P^{(n)}(x,t;\,g) .
\eeq
The function $\rho^{(l)}(x,t;\,g,N)$ represents a contamination coming from the resonances 
with order greater than $N$, to the $l$-th resonance state ($l\le N$).
As already explained, at sufficiently large times, such a contamination should be 
negligible, because the high-energy resonances have small lifetimes and small couplings
to $\phi^{(l)}(x,t=0;\,g,N)$.

We will review the perturbative computation of the matrix $V(g)$,
as well as of its inverse, in the next section, and we will
present some elements of the exact theory in the final
section of the paper.

\section{Perturbative Analysis}
\label{sect4}

By perturbative computation we mean an expansion in powers
of $g$, for $|g| \ll 1$, of all the relevant quantities: wavevectors, 
frequencies, decay widths, mixing matrices, etc.
In order to simplify the formulas and identify
some structure, if any, it is convenient to allow for an ad-hoc
normalization of the resonances.
That is accomplished by defining the "renormalized resonances'' out of the starting (or "bare") ones:
\beq
\xi^{(n)}(x,t;g) \, \equiv \, \frac{ \theta^{(n)}(x,t;g) }{ Z^{(n)}(g) } ,
\eeq
where $Z^{(n)}(g)$ is a function of $n$ and $g$ whose value will be
specified later.  
Eq.(\ref{defVtheta}) is then rewritten as  \cite{secondo}
\beq
\label{defUcsi}
\psi^{(l)}(x,t;g) =  
\sum_{n=1}^{\infty} U_{l n}(g) \, \xi^{(n)}(x,t;g) \, ,
\eeq
where the renormalized mixing matrix $U(g)$ is given, by consistency,
by (we have just multiplied and divided $\xi^{(n)}(x,t;g)$ by the symbol 
$Z^{(n)}(g)$):
\beq
\label{wmrenorm}
\,\,\,\,\,\,\,\,\,\,\,\,\,\, \, \, \, \, \, \, \, \, \, \, \, \,\,\,\,\,
\,\,\,\,\,\,\,\,\,
U(g)_{l,n} =  V(g)_{l,n} \, Z^{(n)}(g)
\, \, \, \, \, \, \, \, \, \, \, \,  \, \, \, \, 
\, \, \, \, \, \, \, \, \, \, \, \, \, \, 
\,\,\,\,\,\,\,
\big({\rm no} \, \, \sum_n\big) .
\eeq
We then insert the small-$g$ expansion for the $k^{(n)}(g)$'s inside
the functions entering the r.h.s. of eq.(\ref{defVtheta}):
\beq
\label{insert2}
k^{(n)}(g) = n - n g + \left( n - i \pi  n^2 \right) g^2 + 
\Big(\frac{4 }{3} \pi ^2 n^3 + 3 i \pi n^2-n\Big) g^3
+ \mathcal{O}\left( g^4  \right) .
\eeq
For the frequencies and widths, we obtain: 
\bea
\omega^{(n)}(g) &=& n^2 \left( 1 - 2g + 3 g^2 \right) \, + \mathcal{O}\left( g^3 \right) \, ;
\\
\label{Gammas}
\Gamma^{(n)}(g) &=&  4 \pi n^3 g^2 
\left(1 \, - \, 4 g \right)\, + \, {\mathcal O}\left( g^4 \right) \, .
\eea
By choosing for instance \cite{secondo}
\beq
Z^{(n)}(g) = 1  + \frac{g}{2}  + \left( \frac{3}{2} i \pi n - \frac{1}{8} \right) g^2  
+  {\mathcal O}\left(g^3\right),
\eeq
$U(g)$ reads in matrix notation, up to second order included:
\beq
\label{finalU2}
U(g_r) = 1 + g_r A + \frac{1}{2} g_r^2 A^2 + i \pi g_r^2 A H 
+ {\mathcal O}\left(g_r^3\right) ,
\eeq
where we have introduced the renormalized coupling
\beq
g_r \equiv g - \frac{1}{2} g^2  +  {\mathcal O}\left(g^3\right) .
\eeq
$A$ is a real antisymmetric matrix with entries
\beq
\label{Aexplicit}
A_{l,n} \equiv (-1)^{l+n} \frac{  2 \, l \, n }{ l^2 - n^2 } 
\, \, \, \, {\rm for} \, \, \, \, \, l \ne n ,
\, \, \, \, \, \, A_{l,l} = 0 
\eeq
and $H$ is the real diagonal matrix with elements
\beq
H_{l,n} \, = \, l \, \delta_{l,n} ,
\eeq
where $\delta_{l,n}=1$ for $l=n$ and zero otherwise is the
Kronecker delta. 
Let us remark that since for large $n$
\beq
\Big| \sin \left[ k^{(n)}(g) x \right] \Big|
\approx \frac{1}{2} e^{\pi g^2 n^2 x} ,
\,\,\,\,\,\,\,\,\,\,
{\rm while}
\,\,\,\,\,\,\,\,
e^{ - 1 / 2 \, \Gamma^{(n)}(g) t } \approx e^{-2 \pi g^2 n^3 t},
\eeq
the series on the r.h.s. of eq.(\ref{defVtheta}) is convergent
for any $t>0$.
In sec.\ref{sect6} we will see that a similar convergence mechanism
also holds in the non-perturbative case (schematically:
$n^2 \to \ln n$ while $n^3 \to n \ln n$).

As already discussed, it is not straightforward to define the inverse
of an infinite matrix, as it not a matter of linear algebra
(the determinant, for example, is not defined).
However, a formal inverse of $U\left(g_r\right)$ exists in perturbation theory
and is given, again up to second order in $g_r$, by:
\beq
\label{finalU^-1}
U^{-1}(g_r) = 1 - g_r A + \frac{1}{2} g_r^2 A^2 - i \pi g_r^2 A H 
+ {\mathcal O}\left(g_r^3\right) .
\eeq
Let us then consider the initial state
\beq
\label{rotateren}
\varphi^{(l)}(x, t = 0;\, g) = \sqrt{\frac{2}{\pi}} \theta(\pi-x) 
\sum_{n=1}^\infty \big(U(g)^{-1}\big)_{l n}  
\, \sin(n x) .
\eeq
The initial wavefunction is naturally decomposed in three
structurally different contributions:
\beq
\varphi^{(l)}(x, t=0;\, g) \, = \,
\alpha^{(l)}(x;\, g) \, + \,
\beta^{(l)}(x;\, g) \, + \, 
\gamma^{(l)}(x;\, g) .
\eeq
The first contribution is obtained by omitting the last 
${\mathcal O}\left(g^2\right)$ term  on the r.h.s. of 
eq.(\ref{finalU^-1}) and reads:
\bea
\label{semplice}
\alpha^{(l)}(x;\, g) & \simeq & \sqrt{\frac{2}{\pi}} \, 
\left( 1 - \frac{g}{2} \right) \, \theta(\pi-x) \,
\sin \left[ l ( 1 - g ) \, x \right] 
\nonumber\\
&\simeq& 
\theta(\pi-x) \, \frac{1}{Z^{(l)}(g)} \, \sqrt{\frac{2}{\pi}} \, 
\sin \big[ k^{(l)}(g) x \big]
\nonumber\\
&\simeq& \theta(\pi-x) \, \xi^{(l)}(x, t=0 ; \, g) \, ,
\eea
i.e. it is basically the renormalized resonance at $t=0$
inside the cavity, continued to zero outside.
The physical interpretation of the above equation is quite
appealing: if you want "diagonal" time evolution, you have to prepare
the initial state with the wavevector $k^{(l)}(g)$ of the
$l$-th resonance, not of the box eigenfunction $k^{(l)}(0)=l$.
The problem is that the above function has a finite $\mathcal{O}(g)$ 
discontinuity at the border of the cavity,
\beq
\lim_{x\to\pi^-} \alpha^{(l)}(x;\, g) \simeq \sqrt{2\pi}(-1)^{l+1} l g ,
\eeq
while the ordinary eigenfunctions or the generalized eigenfunctions 
(resonances) are continuous, so there is a substantial loss of regularity.
Actually, a discontinuity produces an infinite  average kinetic energy. 
Schematically, with
\beq
\psi(x) \, \approx \, \theta(\pi-x)
\,\,\,\,\,\,\, \mathrm{for} \,\,\, x \approx \pi ,
\eeq
we have indeed
\beq
\langle T \rangle \, \approx \,  
\int_{\pi-\varepsilon}^{\pi+\varepsilon} \left( \frac{d \psi}{d x} \right)^2 dx \, = \,
\delta(0) = \infty, 
\eeq
where $0 < \varepsilon \ll 1$.
With the natural regularization coming from the Fourier representation, we have 
in our case:
\beq
\langle T \rangle = 
\int_0^\pi \left(\frac{\partial \alpha^{(l)}(x;\, g)}{\partial x} \right)^2 dx
\, \approx \, 2\pi  l^2 g^2 \, N ,   
\eeq
where we have truncated the sine series to the $N$-th harmonic;
one could choose for example $N \approx 1/g$, obtaining then an 
$\mathcal{O}(g)$ correction to the leading-order kinetic energy.
Furthermore, the average potential energy of the Winter Hamiltonian, 
which involves the operator $\delta(x-\pi)$, is not well defined for a 
wavefunction with a finite discontinuity at $x=\pi$
\footnote{
Formally: 
\beq
\alpha^{(l)}(x;\, g) \in L^2\left(\RR^+\right)
\,\,\,\,\,\,\, \mathrm{but} \,\,\,  \alpha^{(l)}(x;\, g) \notin H^1\left(\RR^+\right) ,
\eeq
where $H^1\left(\RR^+\right)$ is the Sobolev space of the square-integrable
functions on $\RR^+$ with square-integrable (weak) first derivative again 
in $\RR^+$ \cite{brezis}.
}.
That however is not the whole story and an even greater concern is with the 
term $-i \pi g_r^2 A H$. The latter comes from the imaginary term $-i\pi n^2 g^2$ in the
expansion above for $k^{(n)}(g)$ and reads:
\beq
\left( A H \right)_{l,n} = \frac{ (-1)^{l+n+1} \, 2 l n^2 }{ n^2 - l^2 } 
\, \, \, \, \,
{\rm for } \, \, l \ne n \, \, \, \, \, {\rm and} \, \, 0 \, \, {\rm otherwise} \, .
\eeq
Since the latter has a large size for $n\gg l$, it is naturally decomposed as:
\beq
\label{natdecomp}
\left( A H \right)_{l,n} = (-1)^{n+l+1} \, 
\frac{ 2 \, l^3 }{ n^2 - l^2 } + (-1)^{n+l+1} \, 2l . 
\eeq
The second contribution to the initial wavefunction, coming
from the first term on the r.h.s. reads, omitting the trivial factor
$\sqrt{2/\pi} \, \theta(\pi-x)$:  
\beq
\beta^{(l)}(x;\, g) =
- 2 \pi i \, g_r^2 \, l^2 \, 
\ln \left(\cos\frac{ x }{ 2 } \right) \, \sin (l x) + \cdots ,
\eeq
where by the dots we mean less singular terms for $x\to \pi^-$
with respect to the one explicitly written.
Unlike the "renormalized contribution" $\alpha^{(l)}(x;g)$, the above function is continuous 
for $x\to \pi^-$. Actually, it is proportional to the lowest-order term $\sin(l x)$,
with a proportionality constant diverging logarithmically for $x\to\pi^-$.
The best interpretation of this term we could think of, is that of some kind of 
imaginary contribution to the renormalization constant $Z^{(l)}(g)$,
diverging for $x\to \pi^-$, i.e. approaching the border of the cavity
from the inside\footnote{
In general, an imaginary contribution to the (on-shell) renormalization
constant $Z$ of a quantum field signals instability of the
related (massive) particle \cite{mt}.}.

The third contribution to the initial wavefunction is the one
coming from the second term on the r.h.s. of eq.(\ref{natdecomp}) 
and is the most singular one:
\beq
\label{tointepret}
\gamma^{(l)}(x;\, g) = 
i \pi g_r^2 (-1)^{l+1} l \, \tan\left(\frac{ x }{ 2 }\right) + \cdots.
\eeq
The function above is not proportional to the lowest-order wavefunction
$\approx\sin(l x)$ and contains a power divergence 
$\approx 1/(\pi-x)$ for $x\to\pi^-$. 
Let us remark that one cannot even interpret 
eq.(\ref{tointepret}) in a weak sense.
In the latter case one should indeed resum the
involved series as:
\beq
\label{weaksense}
\sum_{n=1}^\infty (-1)^{n+1} \sin (n x) 
= \frac{1}{2} \, {\mathrm P.V.}\tan \frac{x}{2} ,
\eeq
where ${\mathrm P.V.}$ denotes the principal value.
Since the expression on the r.h.s. has then to be multiplied by 
$\theta(\pi-x)$, one ends up with an ill-defined distribution product.
Then $\gamma^{(l)}(x;\, g)$ is not locally integrable and, as a consequence, 
not square integrable. 
The conclusion is that $\gamma^{(l)}(x;\, g)$ is totally unacceptable 
from physics viewpoint.

To summarize, without further input, the best one could do is to 
use the truncation method described in the previous section to
identify operationally the excitations of the system.
One could arbitrarily choose some $N$ and then see the
dependence of the results on the truncation order
to find some stability interval for $N$, if any.
Actually, in the next sections, by means of a global study of the 
function $k=k(g)$, satisfying $b[k(g),g]\equiv 0$ for any $g\in\CC$, we will 
be able to determine the natural truncation order of the perturbative results 
and to compute some non-perturbative effects as well.
The conclusion we will reach is that for fixed $g\ne 0$ the above formulas turn out 
to be applicable for small $N$ only:
\beq
N \, \lsim \, \frac{1}{2\pi |g|} .
\eeq
For larger $N$, one has to use quite different formulas for the
mixing matrix $V(g)$ (or $U(g)$), which have a completely different asymptotic behavior.

\section{The Multivalued Function $h(z)$}
\label{sect5}

By defining
\beq
z \equiv - g ; \, \, \, \, \, \, \, \, w \equiv 2 \pi i \, k ,
\eeq
the problem is, as discussed in the previous section, 
that of inverting the entire function
\beq
f: \CC_w \to \CC_z
\eeq
defined, according to eq.(\ref{bzero}), by
\beq
\label{defg}
z = f(w) \equiv \frac{ e^w-1}{w} 
\eeq
for $w \ne 0$ with $f(0) \equiv 1$.
We want indeed to know the possible wavevectors $k$ as a 
function of the given coupling $g$, i.e. at fixed physics.
Since $f(2\pi i n)=0$ for $n\ne 0$ integer, $f$ is not
injective (one-to-one) and the formal inverse $h$ is
not single-valued. We then proceed to its analysis by 
means of the following steps:
\begin{enumerate}
\item
We identify the branch points of $h$ by means of a local
study of the inverse $f$ \cite{donna};
\item
We perform local expansions of $h$ at various points and
we match them by evaluation on the overlaps of the convergence 
regions (whenever not empty), or by means of proper paths
connecting them;
\item
We cut the complex $z$-plane and then we glue the sheets
in order to finally construct the Riemann surface of $h$.
\end{enumerate}

\subsection{Branch Points}

\begin{figure}[ht]
\begin{center}
\includegraphics[width=0.5\textwidth]{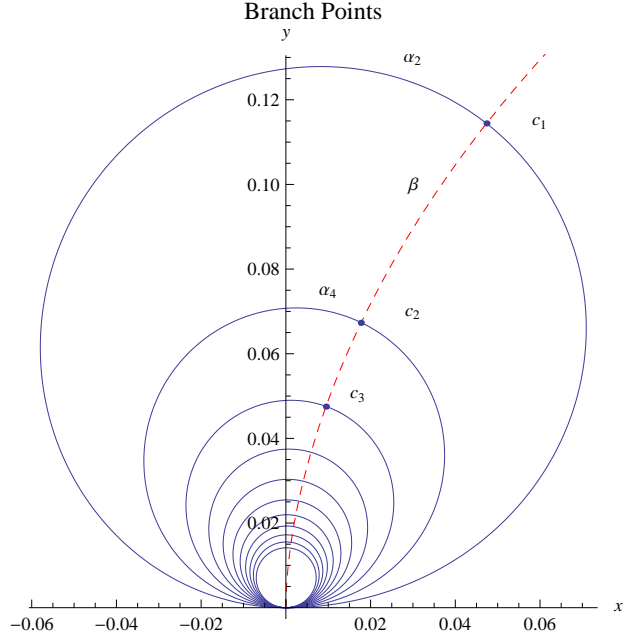}
\footnotesize
\caption{
\label{figura_appB}
\it Branch points $c_n$'s in the first quadrant of the $z$-plane
obtained as intersections of the dashed (red) curve 
$\beta$ with the continuous (blue) curves $\alpha_{2n}$ (see appendix \ref{appB}).}
\end{center}
\end{figure}

When the first derivative of $z=f(w)$ vanishes, let's
say in $w=d_*$, this function is not one-to-one 
in a neighborhood of $d_*$, so the inverse 
\beq
h: \CC_z \to \CC_w ,
\eeq
$w=h(z)$, is multivalued 
in a neighborhood of $c_* = f(d_*)$.
The derivative
\beq
f'(w) = \frac{ (w-1) \, e^w + 1 }{ w^2 }
\eeq
vanishes when $w \ne 0$ is a solution of 
the transcendental equation
\beq
\label{solinw}
e^{-w} = 1 - w \, .
\eeq
By using eq.(\ref{defg}), the above equation can be transformed
into the following equation in $z \ne 1$:
\beq
\label{solinz}
e^{1-1/z} =  z .
\eeq
The order of the branch point can be determined by evaluating the
higher-order derivatives of $f(w)$ in $d_*$.
It holds 
\beq
f''(w) 
= \frac{ \left( w^2 - 2 w + 2 \right) e^w -2}{ w^3 } , 
\eeq
so that
\beq
\left. f''(w) \right|_{ f'(w) = 0 }   = 
\frac{ 1 }{ w ( 1 - w ) } \ne 0 .
\eeq
Since the second derivative does not vanish, 
we have that for $\left|w-d_*\right|\ll 1$
\beq
z-c_* \, \approx \, (w-d_*)^2 
\,\,\,\,\,\,\, 
\Rightarrow
\,\,\,\,\,\,\,
w-d_* \, \approx \, (z-c_*)^{1/2} ,
\eeq
so $w=h(z)$ has a first-order branch point in $z_*$.
Let us call $\{c_n\}_{n\ne 0}$ the set of the solutions
to eq.(\ref{solinz}) different from $z=1$
(which is not a branch point). 
The relation between the images of the branch points $d_n$
and the branch points $c_n=f\left(d_n\right)$ reads:
\beq
c_n \, = \, \frac{1}{1-d_n},
\, \, \, \, \, \, \, \, n\ne 0.
\eeq
As shown in appendix \ref{appB}, all the $c_n$'s lie in the first and
the fourth quadrant of the $z$-plane (see fig.\ref{figura_appB}). 
Let $\{c_n\}_{n>0}$
be the solutions on the first quadrant,
ordered according to decreasing distance 
from the origin,
\beq
\label{ordercn}
|c_1| > |c_2| > |c_3| > \cdots \cdots > 
|c_{n-1}| > |c_n| > |c_{n+1}| > \cdots > 0 ,
\, \, \, \, \, \, \, \, \, \, \, \, \,
\forall n > 1 .
\eeq
\begin{table*}[!htb]
\vspace{-.1in}
\begin{center}
\caption{\it Exact and asymptotic values of the first few
branch points $c_n$'s (of order one) of $h$, 
together with the convergence
radii $R_n$'s of the power expansions of the $h_n(z)$'s around $z=0$ 
and the convergence radii $\rho_n$'s of the semi-integer power
expansions around the $c_n$'s (see main text).}
\vspace{0.1in}
\begin{tabular}{l|c|c|c|c}
\hline
\hline 
\multicolumn{3}{c}{Branch points of $h(z)$ and related quantities}
\\
\hline
$n$ &         $c_n$         &  $c_n^{as}$             & $R_n$           &    $\rho_n$
\\
\hline
$1$ & $ \, 0.0473642 + i \, 0.114414$  & $ 0.0470313 + i \, 0.114685$   & $0.123830$ & 0.0632582
\\
\hline
$2$ & $ \, 0.0177821 + i \, 0.0673565$    & $0.0177176  + i \, 0.067393$  & $0.0696642$ & 0.0214698
\\
\hline
$3$ & $ \, 0.00946886 + i \, 0.0475615$    & $ 0.0094478 + i \, 0.0475706$ & $0.0484949$ & 0.0114077
\\
\hline
\hline
\end{tabular}
\label{tabella1}
\end{center}
\end{table*}
Since eq.(\ref{solinz}) has real coefficients, solutions
come in pairs of complex-conjugated points and therefore
we define
\beq
\label{simmetria}
c_n \, \equiv \, \overline{c_{-n}} \, \, \, \, \, \, \, \, \, \, \, \,
{\rm for} \, \, \, \, \, n < 0 .
\eeq 
Exact values for the first few $c_n$'s are given in the table.
An accurate analytic formula (see appendix \ref{appB}) turns out to be:
\beq
c_n^{as} \, = \, 
\frac{i}
{ \pi(2n+1/2) + i\ln\big[ e \pi (2n+1/2) \big] 
- \ln \big[ e \pi (2n+1/2) \big]/(\pi(2n+1/2))},
\, \, \, \, \, \, \, \, \, \, \, \, \, 
n \ge 1,
\eeq
where ``$\ln$'' is the logarithm of real analysis.
The relative error with respect to the exact value is around
$0.3\%$ for $n=1$ and quickly goes to zero for increasing $n$
(see Table \ref{tabella1}).

The first derivative of $f(w)$ also vanishes for
$u\to-\infty$, with $w=u+iv$ and $u,v\in\RR$.
That would suggest that also
\beq
z = 0 = \lim_{ u \to - \infty } f(u+iv) 
\eeq
is a branch point of $h(z)$. That however is not true
because the origin is not an isolated singularity
being, as we have just seen, an accumulation point 
of the order-one branch points.

\subsection{The Point at Infinity}

Let us now consider the point at infinity.
We make the usual change of variable $\zeta\equiv 1/z$,
so that
\beq
\zeta = \zeta(u+iv) = \frac{ u+i v }{ e^{ u+i v } - 1 } ,
\eeq
and study the limit $\zeta\to 0$.
In that limit, $u\to\infty$ and
\beq
\frac{d\zeta}{dw}(u+iv) \, = \,  
\frac{ (1-u-iv) \, e^{ u+i v } -1 }{ ( e^{ u+i v } - 1)^2 } \, \to \, 0.
\eeq 
The point at infinity is therefore a branch point of $h$.
Evaluating the higher-order derivatives of $\zeta(w)$,
one easily finds that they all vanish for $u\to\infty$,
implying that $z=\infty$ is a branch point of infinite
order for $h(z)$. Indeed, for $u\gg 1$, 
\beq
f(u+iv) = \frac{e^{u+iv}-1}{u+iv} \cong 
\frac{e^{u+iv}}{u+iv}  \simeq \frac{e^{u+iv(1-1/u)}}{u}
\approx \frac{e^{u+iv}}{u},
\eeq
where the last two relations hold for $v\ll u$, for example
$v = \mathcal{O}(1)$.
A sequence of progressively more crude approximations has been
generated above, the last one implying that $z=\infty$ should be a 
logarithmic branch point for $h$. We will see later that $h(z)$ 
actually has an infinite-order branch point at $z=\infty$,
originating from a countable set of order-one branch points at infinity,
which "conspire" to produce a singularity resembling a logarithmic
branch point. 

Since $f(2\pi i n)=0$ for any $n\ne 0$, $f(0)=1$, and $f'(2\pi i n)\ne 0$ 
for any $n$, local inverses $h_n(z)$ do exist because of the inverse function
theorem, satisfying
\beq
h_n(0) = 2 \pi i n \, \, \, \, \, \,\,\,\,\,
{\rm for} \, \, \, n \ne 0;
\, \, \, \, \, \,\,\,\,\,h_0(1) = 0.
\eeq
We will analytically continue the above branches to properly
cut $z$-planes in the next sections. 

\subsection{The Branch $h_0$}

\begin{figure}[ht]
\begin{center}
\includegraphics[width=0.5\textwidth]{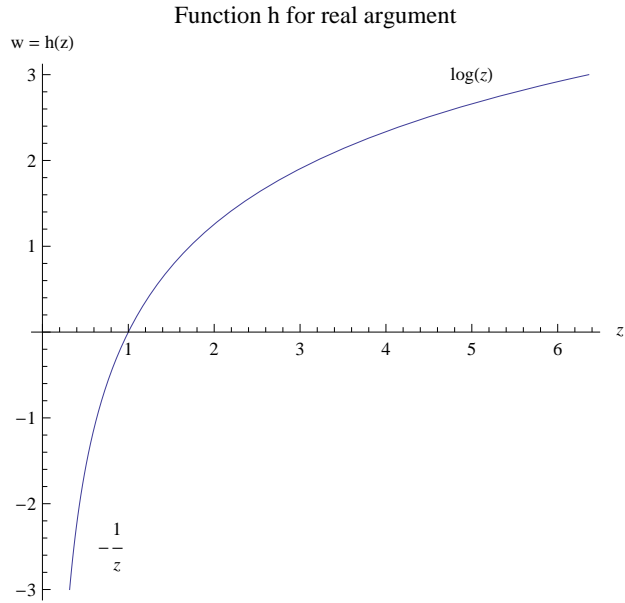}
\footnotesize
\caption{
\label{figura_h}
\it Real plot of the function $h$ for a positive argument, i.e.
of the branch $w=h_0(z)$ for $z\in\RR^+$.
For $z\gg 1$, $h_0(z) \approx \mathrm{log}(z)$, while for
$0<z\ll 1$, $h_0(z) \approx -1/z$.}
\end{center}
\end{figure}

By evaluating both sides of eq.(\ref{relW*h0}) in $z=1$
(see appendix \ref{appA}), one finds that this equation 
relates $h_0$ to the principal branch of the Lambert $W$ function:
\beq
\label{relWh}
h_0(z) = \, - \, \frac{1}{z} \, - \, 
W_* \left( - \frac{ e^{-1/z} }{z} \right) .
\eeq
\begin{figure}[ht]
\begin{center}
\includegraphics[width=0.5\textwidth]{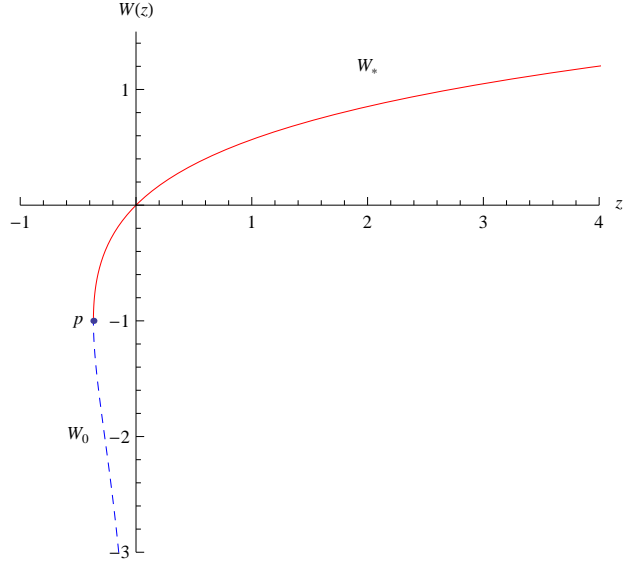}
\footnotesize
\caption{
\label{figura_W}
\it Real plot of the Lambert function $w=W(z)$ for 
real $z \ge -1/e$.
Principal branch $W_*$: continuous (red) line; 
zero branch $W_0$: dashed (blue) line. The two curves 
are separated by the point $p=(-1/e,-1)$, where $z=-1/e$
is the order-one branch point of $W$ and $w=-1$ its image
(see appendix \ref{appA}). 
For $z\gg 1$, $W_*(z)\approx \mathrm{log}(z)$, while for $z<0$,
$|z|\ll 1$, $W_0(z)\approx \mathrm{log}(-z)$.}
\end{center}
\end{figure}
$W_*$ has the following series expansion around the origin
\cite{knuth}:
\beq
W_*(\zeta) = 
\sum_{n=1}^{\infty} (-1)^{n-1} \frac{ n^{n-1}}{n!} \, \zeta^n , 
\eeq
which is convergent for
\beq
|\zeta| \le \frac{1}{e} .
\eeq
The above one is actually the expansion of the bound-state quantum
number $\chi(g)$ for $|g|\ll 1$, i.e. for a tightly-bounded
particle.
Since $\exp(-1/c_n)/c_n = 1/e$, by evaluating $h_0$ in $c_n$, 
we obtain
\beq
h_0(c_n) = \, - \, \frac{1}{c_n} \, - \, 
W_* \left( - \frac{1}{e} \right) .
\eeq
The numerical series appearing on the r.h.s. can be exactly summed 
\cite{knuth}:
\beq
\sum_{n=1}^{\infty} \frac{ n^{n-1} }{n! \, e^n} \, = \, 1 \, ,
\eeq
implying that
\beq
h_0(c_n) = 1 \, - \, \frac{1}{c_n} = d_n ,
\eeq
for any $n \ne 0$.
Therefore we conclude that all the order-one branch points of
$h$ lie on the sheet $S_0$, where $h_0$ is defined 
(see fig.\ref{figura_h} for a real plot).
Since an order-one branch point glues together two different sheets,
we have to find the second sheet containing the branch point $c_n$ for
any given $n\ne 0$. That is accomplished in the next section.

Another relevant expansion for $h_0$ is the one centered in $z=1$,
which is related to the case of the "loosely bounded particle''
discussed before:
\beq
h_0(z) = \sum_{n=1}^\infty k_n (z-1)^n, \,\,\,\,\,\,\,\,\,\,\,\,\,\,\,\,\,\, 
|z-1| < r,
\eeq
where, by Abel's theorem,
\beq
r = \inf_{n\in\ZZ} \left|1-c_n\right| = \left|1-c_1\right| \cong 0.959482 .
\eeq
The first few coefficients explicitly read:
\bea
k_1 &=& + 2 ;
\\
k_2 &=& - \frac{4}{3} ; 
\\
k_3 &=&  + \frac{10}{9} .
\eea

\subsection{The Branches $h_n$ for $n \ne 0$}

By evaluating a large number of terms (up to $10^3$) in the power-expansion
of $h_n(z)$ around $z=0$ for various $n=1,2,3,\cdots$,
\beq 
\label{expand_hn}
h_n(z) = \sum_{k=0}^{\infty} a_k^{(n)} z^k, 
\, \, \, \, \, \, \, \, \, |z| < R_n,
\eeq
and using the standard convergence tests, we find that
\beq
R_n = |c_n| \, \approx \, \frac{1}{ 2 \pi |n| }
\eeq
for any $n \ne 0$.
Let us remark that eq.(\ref{expand_hn}) is actually the perturbative 
expansion of $k^{(n)}(g)$ in powers of $g$ given in eq.(\ref{insert2}) 
in a different notation, so we do not repeat the values of the 
lowest-order coefficients.
Since the $c_n$'s are ordered according to their modulus
(see eq.(\ref{ordercn})),
we conclude that the sheet $S_1$ only has the branch point
$c_1$ (plus an order-one branch point at infinity), $S_2$ has 
only the branch point $c_2$ and so on.
Therefore:
\beq
h_n(c_n) = h_0(c_n) = d_n = 1 - \frac{1}{c_n}, \, \, \, \, \, \, \,\,\,\,
n \ne 0
\eeq
while
\beq
h_k(c_n) \ne d_n \,\,\,\,\, \, \, \, \, \, \, {\rm for} 
\,\,\, \, \, \, k \ne n .
\eeq
The branch point $c_n$ then glues together the sheet $S_n$
to $S_0$ for any $n\ne 0$. 

\subsection{Expansions Around the Branch Points $c_n$'s }

In this section we present an expansion of $h(z)$ around
the branch point $c_n$ for some fixed $n \ne 0$.
Since the latter is an order-one branch point,
the series involves semi-integer powers of 
$z-c_n$.
The expansion is obtained by inverting the Taylor
series of $z=f(w)$ around $d_n$, with $c_n=f(d_n)$,
\beq
z - c_n = \sum_{k=2}^\infty \frac{f^{(k)}(d_n)}{k!} (w-d_n)^k .
\eeq
We replace on the r.h.s. of the above equation 
the semi-integer power (Puiseaux) expansion
\beq
w-d_n = \sum_{l=1}^\infty b_l^{(n)} \, (z-c_n)^{l/2},
\eeq
expand everything in powers of $z-c_n$
and determine the coefficients $b_l^{(n)}$ in such a way that equality
holds order by order in $z-c_n$ \cite{donna}\footnote{
This method is a generalization to the case $f'(d_n) = 0$
of the standard series inversion.}.
For the first few orders, for example, we have the explicit
expressions:
\bea
b_1^{(n)} &=& \frac{1}{\sqrt{a_2^{(n)}}} ;
\\
b_2^{(n)} &=& - \frac{a_3^{(n)}}{2 \left(a_2^{(n)}\right)^2} ;
\\
b_3^{(n)} &=& 
\frac{ 5 \left(a_3^{(n)}\right)^2 - 4 a_2^{(n)} a_4^{(n)} }{8 \left(a_2^{(n)}\right)^3 \sqrt{a_2^{(n)}}} ,
\eea
where we have defined
\beq
a_k^{(n)} \equiv \frac{f^{(k)}(d_n)}{k!} 
\, \, \, \, \, \, \, \, \, \, \, (k \ge 2).
\eeq
By $\sqrt{a_2^{(n)}}$ we mean an arbitrary but fixed complex
square root of $a_2^{(n)}$, for example the branch with
$-\pi < \arg z \le \pi$ ($1^{1/2}=1)$. 
Changing the convention for $\sqrt{a_2^{(n)}}$ is equivalent to 
go from one determination of $(z-c_n)^{1/2}$ to the other.

To summarize, we have the expansion
\beq
\label{rootexp}
h_{n,0}(z) = d_n + \sum_{k=1}^\infty b_k^{(n)} \, (z-c_n)^{k/2}.
\eeq
The double subscript is to remember that the expansion allows
to compute the values of both branches $h_n(z)$ and $h_0(z)$ for
$z$ sufficiently close to $c_n$.
The above series, by a generalization of Abel's theorem, is convergent up 
to the closest singularity to $c_n$ in the $z$-plane,
\beq
|z-c_n| < \rho_n \, = \, 
\min \Big\{ |c_n-c_{n-1}|, \, |c_n-c_{n+1}| \Big\} 
\, = \, |c_n-c_{n+1}|, 
\, \, \, \, \, \, \, \, \, \, \, {\rm for} \, \, \,  n > 1 ,
\eeq
with $\rho_1 = |c_1-c_{2}|$ and $\rho_{-n} = \rho_n$ for $n<0$.
Exact values for the first few $\rho_n$'s are given in the table.
Asymptotically:
\beq
\rho_n \simeq \frac{1}{2\pi \, n(n+1)} ,
\eeq
for $n\gg 1$. Let us remark that $\rho_n > 0$ for any $n \ne 0$, but tends to zero for 
$n \to \pm \infty$.

\subsection{Matching Different Expansions}

In this section, in order to check the analytic-geometric structure of the
function $h$ previously established, we match the different expansions
obtained.

\begin{figure}[ht]
\begin{center}
\includegraphics[width=0.5\textwidth]{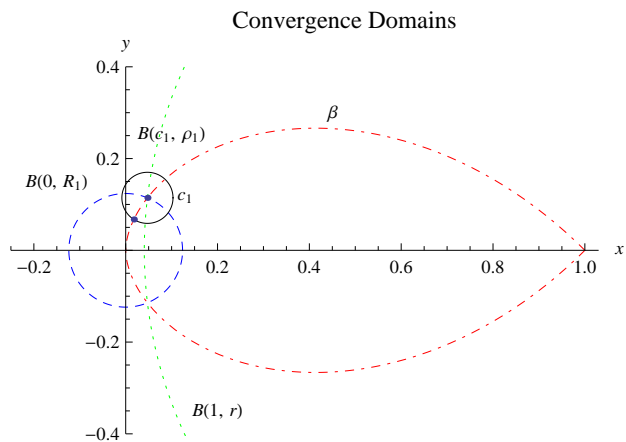}
\footnotesize
\caption{
\label{figura5}
\it Convergence regions of various expansions: 
1) disk with continuous (black) boundary $B(c_1,\rho_1)$: 
semi-integer expansion for both $h_0$ and $h_1$; 
2) ``fish-like'' region with dotted-dashed (red) boundary $\beta$: 
expansion involving exponentials from the Lambert $W$ function for 
the branch $h_0$; 
3) disk with dashed (blue) boundary $B(0,R_1)$: power expansion
centered in the origin for the branch $h_1$;
4) disk with dotted (green) boundary
$B(1,r)$: power expansion of $h_0(z)$ centered in $z=1$.}
\end{center}
\end{figure}

Let us denote with $D\subseteq \CC_z$ the convergence region of the expansion
of $h_0(z)$ coming from the power expansion in zero of $W_*$,
\beq
D \equiv \big\{ z \in \CC \backslash \{0\} : |\exp(-1/z)/z| \le 1/e  \big\} ,
\eeq
with $B\left(0; \, R_n\right)$ the convergence disk of the
expansion of $h_n(z)$ around $0$,
\beq
B\left(0; \, R_n\right) \equiv \big\{ z \in \CC : |z| < R_n  \big\} ,
\eeq
and by $B\left(c_n;\,\rho_n\right)$ the convergence disk of the 
semi-integer expansion of $h_{n,0}(z)$ in $c_n$ derived above
(see fig.\ref{figura5}),
\beq
B\left(c_n;\,\rho_n\right) \equiv \big\{ z \in \CC : |z-c_n| < \rho_n \big\} .
\eeq
We can pick up any point $p$ in the double intersection
\beq
p \in D \cap B\left(c_n;\,\rho_n\right)
\eeq
and fix the sign of $\sqrt{a_2^{(n)}}$ in such a way that,
for example,
\beq
h_{n,0}(p) = + h_0(p) .
\eeq
Because of continuity, the sign is constant on any
connected component of the intersection above.
Since the latter is connected, being $D$ and $B\left(c_n,\rho_n\right)$ 
convex, the sign is unique. 
Then for every point 
\beq
q \in B\left(0;\,R_n\right) \cap B\left(c_n;\,\rho_n\right) ,
\eeq
it must hold
\beq
h_{n,0}(q) = - h_n(q) ,
\eeq
since similar considerations as those above for the sign
also hold in this case.
Other checks of the analytic structure of $h$ can be 
obtained by considering the power-expansion in $z-1$ of
$h_0$ discussed previously, or by taking into account that
\beq
B\left(c_n;\rho_n\right) \cap B\left(c_{n+1},\rho_{n+1}\right) \ne \emptyset , 
\, \, \, \, \, \, \,\,\, n \ge 1.
\eeq
Since $B\left(c_n;\rho_n\right) \cap B\left(c_{n+2},\rho_{n+2}\right) = \emptyset$,
one has to proceed through a chain of disks.
In general, one can analytically continue the function
$w=h(z)$ by moving along paths in the $z$-plane
with the ODE derived in the next section or by means of the 
standard Weierstrass procedure involving a sequence of
overlapping circles \cite{fb}.

\subsection{Cuts}

\begin{figure}[ht]
\begin{center}
\includegraphics[width=0.5\textwidth]{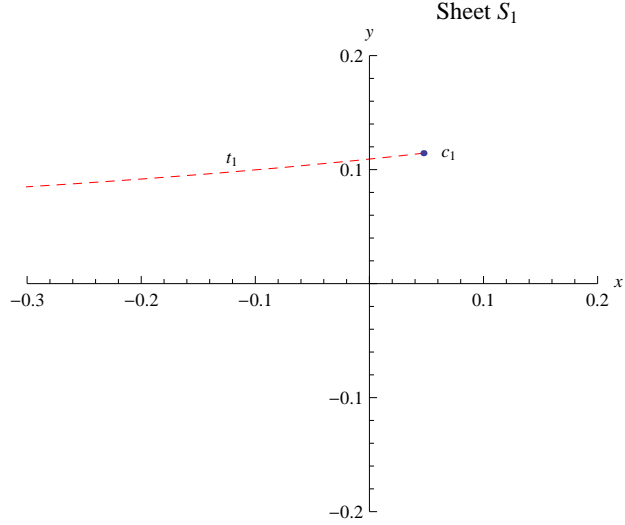}
\footnotesize
\caption{
\label{figura3}
\it Sheet $S_1$ of the branch $h_1$, with
the branch point $c_1$ and the cut $t_1$ from $c_1$ to 
$-\infty$ represented by the dashed (red) curve. }
\end{center}
\end{figure}

\begin{figure}[ht]
\begin{center}
\includegraphics[width=0.5\textwidth]{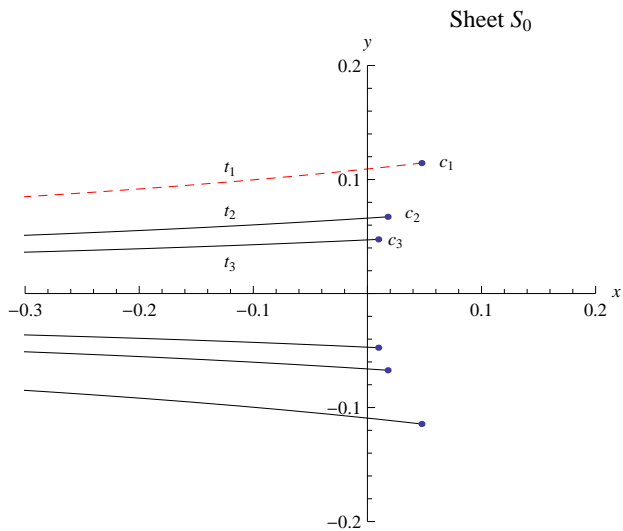}
\footnotesize
\caption{
\label{figura4}
\it Sheet $S_0$ of the branch $h_0$, with
all the order-one branch points $c_n$'s of $h$.
To construct the Riemann surface of $h$,
the borders of the cut $t_1$, represented by the dashed (red) 
curve going from $c_1$ to $-\infty$, have to be glued
diagonally with the corresponding borders of $S_1$ 
(see the previous figure).
All the remaining cuts are plotted as black (continuous) curves
and have to be connected in a similar way to the associated cuts 
in the sheets $S_n$'s.
}
\end{center}
\end{figure}

We cut the $z$-plane of the function $h$
along curves $t_n$ starting from the branch 
points $c_n$ and going to $-\infty$, of the 
(arbitrarily chosen) form
\beq
z_n(t) = a_n - t + i\frac{b_n}{1+t} , 
\, \, \, \, \, \, \, \, \, \, t \ge 0 , 
\eeq
where $c_n = a_n + i b_n$ with $a_n,b_n \in \RR$,
$n\ne 0$.
In fig.\ref{figura3} we plot the $z$-plane cut along $t_1$, 
i.e. the sheet $S_1$ where the branch $h_1$ is defined, 
while in fig.\ref{figura4} we plot the sheet $S_0$, 
where the branch $h_0(z)$ is defined,
having all the cuts $t_n$, $n\ne 0$.
The lifted curves $T_n$ in the $w$-plane are obtained
by integrating numerically the ordinary differential equation
(ODE) for $w = h(z)$, 
\beq
\label{ode}
\frac{dw}{dz} = \frac{w}{z \, w - z + 1} ,
\eeq
along $t_n$.
A singularity of the solution is expected when the denominator, 
a polynomial of second degree in $z$ and $w$, vanishes \cite{hille},
i.e. when:
\beq
w = 1 - \frac{1}{z} .
\eeq
The above equation, combined with the relation between
$w$ and $z$ provided by eq.(\ref{defg}), gives the 
equation for the branch points already obtained
\beq
\label{eqsing}
e^{1-1/z} =  z .
\eeq
By selecting one determination for the function $h$,
let us define
\beq
w_n(t) \equiv w \left(z_n(t)\right) .
\eeq
The ODE for $T_n$ explicitly reads: 
\beq
\label{ode2}
\frac{dw_n}{dt}(t) = \frac{w_n(t)}{z_n(t) \, w_n(t) - z_n(t) + 1} 
\, \frac{dz_n}{dt}(t) .
\eeq
It is clear that an initial condition to the above ODE for $w_n(t)$ cannot 
be provided at $t=0$: solution could not be unique as we would
``begin to move'' starting from a branch point:
\beq
w_n(0) = d_n, \, \, \, \, \, \, z_n(0) = c_n .
\eeq
As we have just shown, $dw/dz$, as given by the r.h.s. of eq.(\ref{ode}), 
formally diverges for $z\to c_n$ and $w\to d_n$. 
We therefore solve the ODE for $t \ge t_0$ with $0 < t_0 \ll 1$
\footnote{
We see here, in a differentiable context, the (algebraic topology) theorem
of the unicity of lifted paths, once an element in the preimage of a
point in the base curve is chosen \cite{fulton}.
}.
The initial condition $w_n(t_0)$ can be derived
by means of the semi-integer expansion (\ref{rootexp})
\beq
w_n\left(t_0\right) = 
d_n + \sum_{k=1}^\infty b_k^{(n)} \, \big( z_n(t_0) - c_n \big)^{k/2} .
\eeq
For each $z_n(t_0)$, there are two $w_n(t_0)$'s
because of the square root terms, as it should
for an order-one branch point.
The two different initial conditions
correspond to the two sides of the cut one
is moving along to.
\begin{figure}[ht]
\begin{center}
\includegraphics[width=0.5\textwidth]{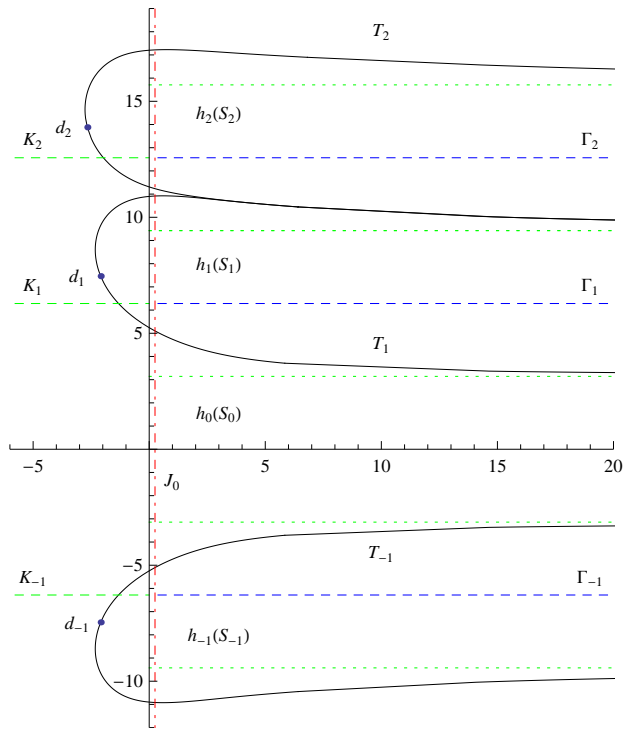}
\footnotesize
\caption{
\label{figura_pwtotal}
\it Images of the sheets $S_n$ under the branches $h_n$ 
in the $w$-plane.
$h_n(S_n)$, $n\ne 0$, is the region inside the continuous
(black) curve $T_n$, with $h_0(S_0)$ the complementary
region in the $w$-plane.
Also shown are simple paths allowing one to visit all the 
Riemann surface $S$ of $h$.
}
\end{center}
\end{figure}
In practice, to obtain a high accuracy, one has to 
compromise about the numerical value of $t_0>0$.
On the one hand, $t_0$ has to be quite close to zero 
in such a way that
$z_n(t_0)-c_n$ is so small that the series expansion
above converges quickly, while on the other hand $t_0$
has to be large enough so that $dw/dz$ is not too big 
in the first iterations of the ODE.

\subsection{Asymptotic Expansions}

For $|\ln_n(z)|\gg 1$, the following asymptotic expansion holds:
\beq
\label{logstart}
h_n(z) = \ln_n(z) + \ln_0\left[\ln_n(z)\right] 
+ \frac{ \ln_0\left[\ln_n(z)\right] }{ \ln_n(z)  }
+ \frac{ 1 }{ z \ln_n(z)  }
+ \mathcal{O}\left[\frac{ 1 }{ \ln_n^2(z)  }\right] ,
\eeq
where by $\ln_n(z)$ we mean the branch of $\ln(z)$ with
argument $\arg_n$, in the range
\beq
\pi (2n-1) < \arg_n(z) < \pi (2n+1) , 
\, \, \, \, \, \, \, \, \, \, \, \,
n\in\ZZ.
\eeq
The dependence on $n$ can be made explicit as:
\beq
\ln_n(z) = \ln_0(z) + 2 \pi i n.
\eeq
For the "external" logarithm, we have (conventionally) taken the
principal branch $\ln_0[\cdots]$; one can actually choose any
other branch, as the difference between different choices cancels
order by order in the expansion, as proved in appendix \ref{appC}.
An important point is that $\ln_n(z)$ becomes large in two
quite different situations: 
\begin{enumerate}
\item
$|z| \gg 1$, any $n$. This case is related to the strong-coupling regime 
of the Winter model;\footnote{
The case $|z|\ll 1$ must be discarded because $z=0$,
as we have seen before, is a non-isolated singularity of $h$, so no 
expansion around the origin of the above form exists.}
\item
$|n|\gg 1$, any $z\ne 0$. This case is related to the high-energy 
excitations of Winter model in the weak-coupling regime, which is 
our main concern.
\end{enumerate}
Actually, in case 1. one can drop the term proportional to $1/z$, 
as it is exponentially small compared to the other ones,
so for $|z|\gg 1$, any $n$, one has:
\beq
h_n(z) = \ln_n(z) + \ln_0\left[\ln_n(z)\right] 
+ \frac{ \ln_0\left[\ln_n(z)\right] }{ \ln_n(z)  }
+ \mathcal{O}\left[\frac{ 1 }{ \ln_n^2(z)  }\right].
\eeq
For physics applications (see next section),
one needs the asymptotic expansion for a real argument.
For $x>0$, the expansion explicitly reads:
\bea
h_n(x) &=& \ln(x) + 2 \pi i n + \ln_0\left[\ln(x) + 2\pi i n\right] 
+ \frac{ \ln_0\left[\ln(x) + 2\pi i n\right] }{ \ln(x) + 2\pi i n  } +
\nonumber\\
&& \, \, \, \, \, \, \, \, \, \, \, \, \, \, \, \, \, \, \, \, \, \, 
\, \, \, \, \, \, \,\,\,\,\,\,\, 
+ \frac{ 1 }{ x \big[ \ln(x) + 2 \pi i n  \big] }
+ \mathcal{O}\left[\frac{ 1 }{ \ln_n^2(x) }\right] ,
\eea
where $\ln(x)$ is the real logarithm of $x>0$.
As in the general case above, for $x\gg 1$, any $n$, one
drops the power-suppressed term proportional to $1/x$.
The argument $2\pi n$ entering the above formula
can be directly obtained by considering the curve in the $w$-plane
(see the dashed (blue) lines to the right of the imaginary
axis in fig.\ref{figura_pwtotal}):
\beq
\Gamma_n:
\, \, \, 
w = w_n(u) = u + 2 \pi i n, \, \, \, \, \, u \ge 0.
\eeq
For $n\ne 0$, by taking the formal inverse of $f(w_n(0))= 0$, 
we obtain $h(0)=w_n(0)= 2\pi i n$,
implying that we are on the branch $n$ at 
the initial point of $\Gamma_n$: $h=h_n$.
As shown previously, for $u\gg 1$, 
\beq
f(w_n(u)) \approx \frac{ e^u }{ u } .
\eeq
We now take the inverse again, of the above formula.
Since the curve $\Gamma_n$ does not cross the preimage
$T_n$ of the cut $t_n$ in $S_n$, it is entirely contained
in $h_n(S_n)$. 
By continuity therefore the inverse is again $h_n$
and $u+2\pi i n \approx h_n(e^u/u)$.
With $x = e^u/u$, we then obtain $h_n(x) \approx \ln(x)+2\pi i n$.
For $n=0$ the argument is similar: one takes the inverse
of $f(w_0(0))=1$ and gets the branch $h_0$; again,
the curve $\Gamma_0$ does not cross any of the curves
$T_n$, so it is entirely contained in $h_0(S_0)$. 
For
\beq
x > 0 \, \, \, \, \, \, \, \,\,\,\, 
{\rm and} \, \, \, \,\,\,\, n \gg \frac{1}{2\pi x},
\eeq 
the desired expansion reads:
\bea
\label{required1}
h_n(x) &=& 2\pi i \left( n + \frac{1}{4} \right) + 
\ln\left[ 2 \pi \left( n + \frac{1}{4} \right) x \right] 
- i \, \frac{ \ln\Big[ 2\pi(n+1/4) x \Big] }{ 2\pi(n+1/4) } +
\nonumber\\
&& \, \, \, \, \, \, \, \, \, \, \,\, \, \, \, \, \,\,\,\,\, 
- \frac{ i }{ 2\pi(n+1/4) \, x }
+\mathcal{O}\left\{  \frac{ 1 }{ \left[2\pi(n+1/4)\right]^2 } \right\} .
\eea
For $n<0$, one just takes the complex conjugate as:
\beq
h_n(x) \, = \, \overline{h_{-n}(x)}.
\eeq
For $x<0$, the logarithmic expansion on the r.h.s. of eq.(\ref{logstart}) 
takes the form:
\bea
\label{numeric}
h_n(x) &=& \ln|x| + i \pi (2n-1)+ \ln_0\left[\ln|x| + i \pi (2n-1)\right] 
+ \frac{ \ln_0\big[\ln|x| + i \pi (2n-1)\big] }{ \ln|x| + i \pi (2n-1)  } +
\nonumber\\  
&& \, \, \, \, \, \, \, \, \, \, \, \, \,\,\,\,\,\,\,\,\,\,\,\,\,\,\,\,\,\,
+ \frac{ 1 }{ x \big[ \ln|x| + i \pi (2n-1) \big] }
+ \mathcal{O}\left[\frac{ 1 }{ \ln_n^2(x) }\right] .
\eea
As in the previous cases, for $x \ll - 1$ and $n\ge 1$, 
one drops the power-suppressed term.
The determination of the argument $\pi (2n-1)$ in eq.(\ref{numeric})
has been obtained by integrating numerically 
eq.(\ref{ode}) along the negative axis, i.e. for $z=-t$ 
with $t\ge 0$ with the initial condition $w(0)=2\pi i n$.
Let us remark that one cannot evaluate $h_0(x)$ for
$x<0$ by integrating the ODE for $dw/dz$ starting
from $x=0$ because, as we have shown, the origin is a 
non-isolated singularity for this branch and therefore 
cannot be used as initial condition.
One cannot even start, for example, from $x=1$ ($h_0(1)=0$),
because one cannot pass through the origin either.
Actually, in the latter case, the following instability occurs.
By departing arbitrarily small from the real axis, a branch cut
$t_n$ of some $c_n$ is necessarily hit, because the $c_n$'s accumulate 
at the origin.
As a consequence, one leaves the sheet $S_0$ and enters the sheet $S_n$, 
for some large $n$, and presumably remains in that sheet, as $S_n$ has 
$c_n$ as its only branch point (apart from the point at infinity).
The consequence is that there is a not a well-defined
value of $h_0(x)$ for $x<0$ as far as "physics''
(which deals with $x\in\RR$ only) is concerned.  
One has instead to move along great half-circles, starting from
positive arguments, ending up with a sign depending 
on the contour orientation.
For
\beq
x<0 \,\,\,\,\,\,\,\,\, {\rm and} \,\,\,\,\,\, n \gg \frac{1}{2\pi |x|},
\eeq 
the required expansion explicitly reads:
\bea
\label{required2}
h_n(x) &=& 2\pi i \left( n - \frac{1}{4} \right) 
+ \ln\left[ 2 \pi \left( n - \frac{1}{4} \right) |x| \right] 
- i \, \frac{ \ln\Big[ 2\pi(n-1/4) |x| \Big] }{ 2\pi(n-1/4) } +
\nonumber\\
&& \, \, \, \, \, \, \, \, \, \, \, \, \,\,\,\,\,\,\, \,\,\,\,\,
- \frac{ i }{ 2 \pi(n-1/4) \, x }
\, + \, 
\mathcal{O}\left\{ \frac{1 } { \left[2\pi(n-1/4)\right]^2 } \right\} . 
\eea
Because of the symmetry between eqs.(\ref{required1}) and (\ref{required2}),
one can write both equations in a unique form. For
\beq
x \in \RR \backslash \{0\} ,
\,\,\,\,\,\,\,\, n \gg \frac{1}{2 \pi |x|} ,
\eeq
it holds:
\bea
\label{requiredfinal}
h_n(x) &=& 2\pi i \left( n + \frac{\mathrm{sign}(x)}{4} \right) + 
\ln\left[ 2 \pi \left( n + \frac{\mathrm{sign}(x)}{4} \right) |x| \right] 
- i \, \frac{ \ln\Big[ 2\pi(n + \mathrm{sign}(x)/4) |x| \Big] }{ 2\pi(n + \mathrm{sign}(x)/4) } +
\nonumber\\
&& \, \, \, \, \, \, \, \, \, \, \,\, \, \, \, \, \,\,\,\,\, 
- \frac{ i }{ 2\pi(n + \mathrm{sign}(x)/4) \, x }
+\mathcal{O}\left\{  \frac{ 1 }{ \left[2\pi(n + \mathrm{sign}(x)/4)\right]^2 } \right\} ,
\eea
where $\mathrm{sign}(x)=1$ for $x>0$ and $-1$ otherwise is the function, 
which returns the sign of $x$.

\subsection{Riemann Surface}

By taking into account the results of the previous
sections, one finds that the Riemann surface $S$ of $h$ is 
constructed by gluing diagonally the borders of the cut $t_n$
from $c_n$ to $-\infty$ in $S_n$ to the corresponding ones
in $S_0$, for any $n \ne 0$\footnote{
That is exactly what we do with the $z^{1/2}$ function.}.
Topologically, $S$ is a plane, as it is usually the case with
non-compact Riemann surfaces \cite{springer,forster}.
The images of the branches $h_n(S_n)$ in the $w$-plane,
the total space of the (branched) covering realized by $f$, 
are shown in fig.\ref{figura_pwtotal}.

\begin{figure}[ht]
\begin{center}
\includegraphics[width=0.5\textwidth]{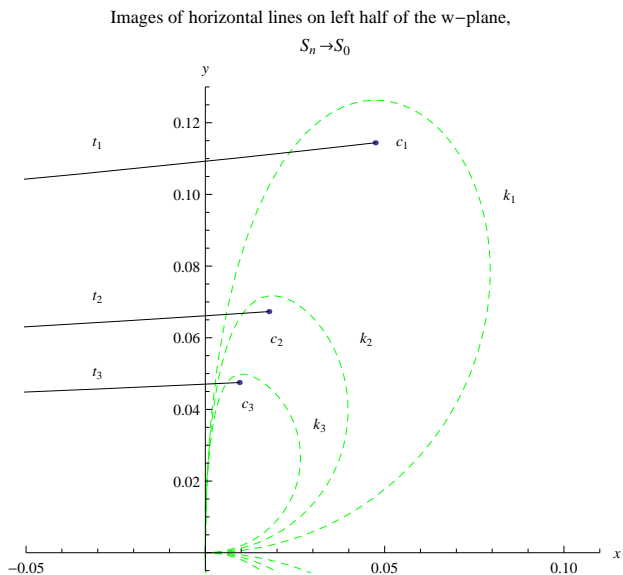}
\footnotesize
\caption{
\label{figura_cross}
\it Image curves $k_n=f(K_n)$, in the $z$-plane, of the horizontal dashed (green) lines
$K_n$ on the left side of the $w$-plane.
By traveling clockwise along $k_n$ (which starts in $z=0$ and tends to $z=0$ for $u\to-\infty$),
one goes from $S_n$ to $S_0$ by crossing the cut $t_n$, painted as a continuous (black) curve.}
\end{center}
\end{figure}

\begin{figure}[ht]
\begin{center}
\includegraphics[width=0.5\textwidth]{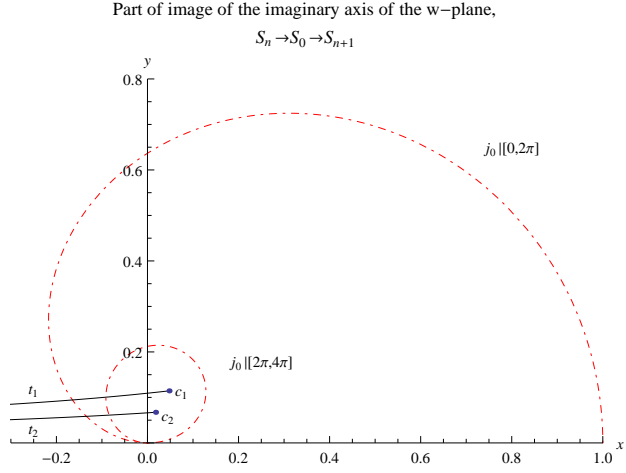}
\footnotesize
\caption{
\label{figura_hard}
\it Part of the image curve $j_0=f(J_0)$, in the $z$-plane, of the
imaginary axis of the $w$-plane, represented by the vertical 
dotted-dashed (red) line $J_0$. 
Traveling counterclockwise along the restriction of $j_0$ to the interval 
$[0,2\pi]$ (starting in $z=1$ and ending in $z=0$), 
one goes from $S_0$ to $S_1$ by crossing the cut $t_1$.
Traveling counterclockwise along $j_0|[2\pi,4\pi]$ (starting and ending at the origin), 
one goes from $S_1$ to $S_0$ by crossing $t_1$ and then from $S_0$ to $S_2$ by
crossing $t_2$.}
\end{center}
\end{figure}

One can visit the entire Riemann surface $S$ by moving along the 
following simple paths, which are also shown in fig.\ref{figura_pwtotal}.
As we have already seen, the dashed (blue) horizontal line on
the right half-plane of $w$,
\beq
\Gamma_n: \,\,\,  w = w_n(u) = u + 2 \pi i n,  \,\,\,\,\,\, u \ge 0,
\eeq
does not cross any continuous (black) curve $T_n$, passing through $d_n$, 
which is the preimage under $h$ of the cut $t_n$.
$\Gamma_n$ is therefore completely contained in $h_n(S_n)$.
Let us remark that the point $d_n$ divides the curve $T_n$ into two arcs,
each of which is mapped by $f$ onto $t_n$\footnote{
It's like with the function $w=z^{1/2}$, where the origin of the $z$-plane
divides any line passing through it into two half-lines, each of which is mapped onto
the same half-line of the $w$-plane.}. 
As already discussed, these "intra-sheet'' paths $\Gamma_n$ allow one to connect 
the power expansion of $h_n(z)$ in $z=0$ with the logarithmic expansion at $z=\infty$. 
The dashed horizontal (green) line on the left half of the $w$ plane, 
\beq
K_n: \,\,\, w = w_n(u) = u + 2\pi i n,  \,\,\,\,\,\, u \le 0, \,\,\, n\ne 0 ,
\eeq
crosses the continuous (black) curve $T_n$ once and therefore allows one 
to go from the sheet $n$ to the sheet $0$ (or viceversa):
\beq
K_n: \, \, \, \, S_n \leftrightarrow S_0 .
\eeq 
The image curves $k_n=f(K_n)$ in the base space $z$ are shown still dashed 
(in green) in fig.\ref{figura_cross}.
The dotted-dashed (red) vertical line,
\beq
J_k: \,\,\, w = w_k(v) = k + i v, \,\,\,\,\,\, 
v\in\RR,  
\eeq
where $k$ is a real constant, crosses all the $T_n$'s 
for $k \gsim {\rm Re} \, d_1$, and allows one to go from $S_n$ to $S_{n+1}$ via $S_0$:
\beq
J_k: \,\,\, S_n \leftrightarrow S_0 \leftrightarrow S_{n+1}, \,\,\,\,\,\, (n\ne 0; \, k \gsim {\rm Re} \, d_1).
\eeq
For $k \gg 1$, the image curve in the base space $j_k=f(J_k)$ is basically 
a great spiral winding an infinite number of times around the point 
$z=\infty$, which we have proven to be an infinite-order branch point 
for $h$.
For $k \ll {\rm Re} \, d_1$ only some sheets are visited; in this case one can still 
go from any $S_n$ to $S_{n+1}$ by passing though $S_0$ via the dashed (green) lines.
$J_0$ coincides with the imaginary axis of the $w$-plane,
its image in the base space $j_0=f(J_0)$
contains the curves $\alpha_n$'s considered
in appendix \ref{appB} (see eq.(\ref{coincide}))
and passes through the origin $z=0$ an infinite number of times.
Closed curves are obtained in this case and all the multivaluedness 
of $h$ is seen in a very clean way.
For $n\ne 0$, one has indeed $f(w_0(n))=0$, implying 
$h_n(0)= 2 \pi i n = w_0(n)$, while $f(w_0(0))=1$ implies 
$h_0(1)=0=w_0(0)$.
A portion of the image curve $j_0=f(J_0)$ in the base space is shown, 
again dotted-dashed (in red), in fig.\ref{figura_hard}

\section{Non-Perturbative Analysis}
\label{sect6}

This is the central section of the paper,
in which the limits of perturbation theory are
rigorously established, non-perturbative results for 
various observables are obtained and compared with the 
perturbative formulas.
The perturbative results for frequencies, 
widths and mixing matrix elements of resonances
have all been obtained by inserting into the exact expressions 
the power expansions for $k^{(n)}(g)=n-ng+\cdots$,
which converge only for
\beq
|g| \, \le \, R_n \, \approx \, \frac{ 1 }{ 2 \pi n } .
\eeq
Therefore, if we consider a fixed resonance (i.e. if we fix $n$),
its properties can be analyzed with arbitrary accuracy by means
of the perturbative formulas for small enough $|g|$.
Actually, we are interested in varying $n$ within a specified
model, i.e. at fixed $g$; by inverting the above inequality,
the following upper bound on $n$ is obtained:
\beq
n \, \lsim \, \frac{ 1 }{ 2 \pi |g| } .
\eeq
Therefore at fixed $g\ll 1$ ("fixed physics"), only the resonances
satisfying the above inequality can be described by means of
perturbation theory.

In order to obtain non-perturbative results, we basically
insert in the general formulas for the observables, the
large-$n$ expansion previously obtained, which we repeat 
here for easy of reference:
\bea
\label{requiredfinalh}
h_n(-g) &=& 2\pi i \left( n - \frac{\mathrm{sign}(g)}{4} \right) + 
\ln\left[ 2 \pi \left( n - \frac{\mathrm{sign}(g)}{4} \right) |g| \right] 
- i \, \frac{ \ln\Big[ 2\pi(n - \mathrm{sign}(g)/4) |g| \Big] }{ 2\pi(n - \mathrm{sign}(g)/4) } +
\nonumber\\
&& \, \, \, \, \, \, \, \, \, \, \,\, \, \, \, \, \,\,\,\,\, 
+ \frac{ i }{ 2\pi(n - \mathrm{sign}(g)/4) \, g }
+\mathcal{O}\left\{  \frac{ 1 }{ \left[2\pi(n - \mathrm{sign}(g)/4)\right]^2 } \right\} ,
\eea
where $\mathrm{sign}(g)$ is the function which returns the sign of $g$.
The basic equation, which replaces the perturbative formula
given in eq.(\ref{insert2}) valid for small $n$, reads:
\bea
\label{insert3}
k^{(n)}(g) &=& \frac{h_n(-g)}{2\pi i} 
\nonumber\\
&=&  n - \frac{\mathrm{sign}(g)}{4} 
- \frac{ i }{ 2 \pi } \, \ln\left\{ 2 \pi |g| \left[ n - \frac{\mathrm{sign}(g)}{4} \right]  \right\}
- \, \frac{ \ln\Big\{ 2 \pi |g| \, \big[n-\mathrm{sign}(g)/4\big] \Big\} }{ 4 \pi^2 \big[ n-\mathrm{sign}(g)/4 \big] } +
\nonumber\\
&& \,\,\,\,\,\,\,\,
+ \frac{ 1 }{ 4 \pi^2 \, g \, \big[ n-\mathrm{sign}(g)/4 \big] } \, + \, 
\mathcal{O}\left\{ \frac{1 }{\left[ 2\pi (n-\mathrm{sign}(g)/4)\right]^2 } \right\} .
\eea
The expansion begins with $n$ in both cases but,
while the perturbative expansion is an expansion in powers of $g$,
the non-perturbative one is basically an expansion in powers of $1/n$.
For 
\beq
0 < |g| \ll 1
\,\,\,\,\,\,\,\,\,\, {\rm and} \,\,\,\,\,\,\,
n \gg \frac{1}{2\pi |g|},
\eeq 
the frequencies and the widths take the form:
\bea
\omega^{(n)}(g) &=&  n^2 - \frac{\mathrm{sign}(g)}{2} n + \mathcal{O}(1) ,
\\
\label{larghezzeNP}
\Gamma^{(n)}(g) &=& \frac{2}{\pi} \, n \ln\big[ 2 \pi |g| n \big] + \mathcal{O}(1) .
\eea
By comparing with the perturbative results obtained
previously, we see that the frequencies have a similar
dependence on $n$ to the free case,
while the widths grow much less with increasing $n$.

Since we do not manipulate perturbative formulas in this
section, there is no more convenience in introducing
renormalization constants and renormalized mixing matrices 
for the resonances, so we switch back to the original 
(unrenormalized) mixing matrix $V(g)$ and resonances $\theta_n$.
In terms of the function $h$, the entries of the mixing matrix $V(g)$
read for large $n$:
\bea
V(g)_{l,n} &=&       
 (-1)^{l+1} \, 2 l g \, \frac{ k^{(n)}(g) } { k^{(n)}(g)^2 - l^2 } \, 
\frac{ e^{ 1/2 \, h_n(-g) } }{ e^{ h_n(-g) } + g } 
\\
&=&  (-1)^{l+1} \, 2 l g \, \frac{ k^{(n)}(g) } { k^{(n)}(g)^2 - l^2 } \, 
e^{-1/2 \, h_n(-g)}
\left[  
1 - g e^{- h_n(-g)} + g^2 e^{- 2 h_n(-g) } + \cdots
\right] .
\nonumber
\eea
By inserting the previous expansion for $h_n(-g)$,
the matrix elements take the explicit form:
\beq
V(g)_{l,n} = \sqrt{ \frac{2}{\pi} } \, \mathrm{sign}(g) \, e^{ i \pi / 4 \, \mathrm{sign}(g) } 
\, \frac{ (-1)^{l+n+1} l \sqrt{ |g| \, n } }{ n^2 - i \, n \log\big[2\pi |g| n\big] /\pi - \mathrm{sign}(g) \, n / 2 - l^2 }   
\left[  
1 + \mathcal{O}\left( \frac{ 1 }{ n } \right)
\right] \, .
\eeq
A few comments are in order:
1) higher-order corrections in $1/n$ can be computed in a
straightforward way; 
2) the alternating sign $(-1)^n$ is present both
in the small-$n$ and large-$n$ expansions
of $V(g)_{l,n}$: it is related to the logarithmic
behavior of the function $h$ at infinity.
As we have seen in the previous
section, this factor is also present in the
inverse of the perturbative expression for $V(g)$,
i.e. in $V_{\rm pt}(g)^{-1}$.
We expect a similar phenomenon to occur also in 
the exact inverse.
Indeed, by looking for example at eq.(\ref{weaksense}), 
one notices that the alternating sign $(-1)^n$ produces constructive 
interference  of the harmonics entering the sine Fourier transform around
$x=\pi$, which is the region where non-trivial dynamics occurs.
For $n\gg l$ with fixed $l$, i.e. along a row 
well to the right to the main diagonal,
\beq
V(g)_{l,n} \simeq	 
\sqrt{ \frac{2}{\pi} } \, \mathrm{sign}(g) \, e^{ i \pi / 4 \, \mathrm{sign}(g) } \,  \, \sqrt{|g|} \,
\, \frac{ (-1)^{l+n+1} l }{ n^{ 3/2 } },
\,\,\,\,\,\,\,\,\,\,\,\,\,\,\,\,\,\,\,\,\, n \gg l .
\eeq
Therefore, for large $n$ there is a rather slow power-law decay, with 
the alternating sign $(-1)^n$.
The above behavior has to be compared with the perturbative one: as
we have seen, $V(g)_{l,n} \approx 1/n$ to $\mathcal{O}(g)$, with
convergence properties deteriorating in second order:
$V(g)_{l,n} \approx 1$ to $\mathcal{O}\left(g^2\right)$.
For $l \gg n$, i.e. along a column well below the main diagonal:
\beq
V(g)_{l,n} \simeq \sqrt{ \frac{2}{\pi} } \, \mathrm{sign}(g) \, e^{ i \pi / 4 \, \mathrm{sign}(g) } 
\, \frac{ (-1)^{l+n} \, \sqrt{ |g| \, n } }{ l } ,
\,\,\,\,\,\,\,\,\,\,\,\,\,\,\,\,\, l \gg n .  
\eeq
A quite slow inverse-power decay with $l$ occurs, with an alternating sign,
exactly as in the perturbative case.
Finally, on the diagonal, one explicitly has:
\beq
V(g)_{n,n} \, = \, e^{ - i \pi / 4 \, \mathrm{sign}(g) } \, 
\frac{ \sqrt{ 2 \pi |g| n } }{ \ln \big(2 \pi |g| n \big) - i \pi \mathrm{sign}(g)/2}
\left[  
1 + \mathcal{O}\left( \frac{ 1 }{ n } \right)
\right], 
\,\,\,\,\,\,\,\,\,\,\,\,\,\,\,\, l=n .
\eeq
Therefore, for large $n$ a square-root divergence basically occurs. 
In perturbation theory we found instead
$V_{n,n} \approx 1$ to $\mathcal{O}(g)$ and
$V_{n,n} \approx n$ to $\mathcal{O}\left(g^2\right)$
(to see that just invert eq.(\ref{wmrenorm}) with respect to $V(g)$ 
or see \cite{secondo}).

A numerical analysis (involving the inversion of matrices
of various order up to $2\times 10^4$) shows that
\beq
\left| \left(V(g)^{-1}\right)_{l,n} \right| \, \approx \, \frac{1}{n};
\,\,\,\,\,\,\,\,\,\,\,\,
\arg\left[ \left(V(g)^{-1}\right)_{l,n+1} \right] 
- \arg\left[ \left(V(g)^{-1}\right)_{l,n} \right] \, \approx \, \pi 
\,\,\,\,\,\,\,\,\,\,\, \mathrm{for} \,\,\,\, n \gg l . 
\eeq
The first relation implies that it is not possible to excite exactly one resonance 
at a time by means of a sensible (i.e. belonging to $H^1\left(\RR^+\right)$) 
initial wavefunction and one has necessarily to use the approximate (truncated) 
scheme.
Note that
\beq
\theta(\pi-x) \sum_{n=1}^\infty \frac{(-1)^{n+1}}{n} \sin(n x) \, = \,  
\theta(\pi-x) \frac{x}{2} .
\eeq
We can summarize the above findings by saying that an exact treatment 
of $V(g)^{-1}$ only partially regularizes the perturbative results.
The inverse matrix elements $\left(V(g)^{-1}\right)_{l,n}$ in the 
non-perturbative case decay faster for $n\to\infty$ than in the perturbative 
case, giving rise to less singular initial data.

\section{Conclusions}
\label{sect7}

The resonances of the Winter model in the weak coupling domain
$0< |g| \ll 1$ --- which is the most interesting one --- are subjected 
to two different regimes.
The first one is the perturbative one, accurately describing
the dynamical properties of the $n$-th resonance up to
\beq
n \, \lsim \, \frac{1}{2 \pi |g|} .
\eeq
The second regime is non-perturbative in character and
describes the dynamics of the $n$-th resonance for
\beq
n \, \gg \, \frac{1}{2 \pi |g|} .
\eeq
Actually, there is also an intermediate region,
\beq
n \, \approx \, \frac{1}{2 \pi |g|} ,
\eeq
which, contrary to the above ones, cannot be described by simple 
analytic formulas.
In physical terms, we may say that perturbation theory accurately describes 
the low-energy excitations of the model, while it completely 
fails to describe the high-energy ones ($|g|\ll 1$ always).
For the decay widths of resonances, for example, we found:
\beq
\Gamma^{(n)}(g) \, \simeq \, 
\left\{
\begin{array}{cc}
4 \pi g^2 n^3   \, + \, \mathcal{O}(g^3) 
\, \, \, \, \, \, \, \, \, \, \, \, \, \, \, \, \,\,\,\,\,\,\,\,\,
{\rm for} \, \, n \lsim 1/(2\pi |g|) \, ;
\\
2/ \pi \, n \ln \big( 2 \pi |g| n \big) 
\, + \, \mathcal{O}(1) \, \, \, \, \, \, {\rm for} \, \, 
 n \gg 1/(2 \pi |g|) \, .
\end{array}
\right.
\eeq
The growth of the widths with $n$ is strongly reduced for large $n$,
\beq
n^3 \, \to \, n \ln n,
\eeq
so that the extrapolation of the perturbative formula above in the
high-energy region would produce completely wrong results.
We have also provided explicit non-perturbative expressions for some 
physically relevant quantities, such as wavevectors, frequencies and mixing-matrix 
entries of resonances, which replace the perturbative formulas previously obtained 
\cite{primo,secondo}.
We have shown that it is not possible to construct in perturbation theory 
initial wavefunctions exciting exactly one resonance at a time.
By inverting numerically the exact matrix $V(g)$ truncated up to
order $N\approx 2 \times 10^4$, we have found that that is not even 
possible to construct such a wavefunction in the exact theory. 

The rigorous limitations of perturbation theory, 
as well as the non-perturbative results we found, 
are all based on the analytic and geometric study 
of the multivalued function 
\beq
w \, = \, h(z) ,
\eeq
which is the inverse of the entire function
\beq
z \, = \, \frac{ e^{w} - 1 }{w} ,
\eeq
where $w \equiv 2\pi i k$ and $z \equiv -g$.
It is necessary to complexify the 
coupling of the model $z$,
i.e. to leave the physical domain $z\in\RR$, in
order to go consistently beyond perturbation theory.
The branch $h_n(z)$, $n\ne 0$, of the function $h(z)$
determines the wavevector of the
$n$-th resonance and therefore its frequency, width, etc.
A knowledge of all these branches is needed to determine
the mixing properties of the resonances.
The branch $h_0(z)$ controls instead the properties
of the bound state, which occurs in the spectrum for 
$0<z<1$.
Each branch $h_n(z)$, $n\ne 0$, is defined on
a cut copy of the plane of the complex variable $z$,
let's call it $S_n$, and has order-one branch 
points in 
\beq
c_n \, \approx \, \frac{i}{2\pi n} 
\eeq
and at infinity.
There is a remarkable connection between physics and mathematics: 
each resonance or bound state of the Winter model is associated to 
a specific sheet of the Riemann surface of the multivalued function
$w=h(z)$, which is the zero locus of the coefficient $b(k,g)$ 
multiplying the backward wave $\exp(-i k x)$
in the continuous-spectrum eigenfunctions.
The transcendental equation satisfied by the $c_n$'s reads
\beq
e^{-1/z} \, = \, \frac{z}{e}
\eeq 
and has been derived in different ways as: 1) the zero locus
of the first derivative of the inverse function of $w=h(z)$;
2) the singularity set of the ordinary differential
equation for $dw/dz$; 3) the preimage of the order-one
branch point $-1/e$ of the Lambert $W$ function under the function
$q(\zeta) = -\exp(-1/\zeta)/\zeta$.
Qualitative properties of the $c_n$'s have been 
determined, together with explicit (approximate) analytic 
formulas. 
The distance of $c_n$ from the origin
determines the convergence radius $R_n$ of the perturbative
expansions (power series in $z$) for the $n$-th resonance:
\beq
R_n \, = \, |c_n| \, \approx \, \frac{1}{2\pi n} .
\eeq
At variance with respect to all the other sheets, 
the "bound-state" sheet $S_0$, where the branch $h_0(z)$ is 
defined, contains has all the order-one branch points.
The fact  that
\beq
\label{heavylim}
R_n \to 0 \,\,\,\,\,\,\,\,\, {\rm for} \,\,\,\,\,\, n \to \pm \infty
\eeq
has various physical consequences.
The first one is that high-energy resonances ($n\gg 1$) admit a 
perturbative description for smaller $|g|$ compared with
the lower-energy ones, in agreement with physical intuition. 
The second consequence is that the point $g=0$ is a non-isolated singularity
in the sheet $S_0$, being an accumulation point of branch points.
Since $g=0$ is a non-analyticity point, no convergent perturbative
expansion for bound-state quantities, such as the energy,
the wavefunction, etc. exists.
We have however provided specific expansions for the bound-state energy:
a power expansion in $g+1$, which is convergent for $|g|$ not too 
small, as well as an expansion for $0<|g|\ll 1$ which involves the 
functions (non analytic in the origin) $1/g^n $ and $e^{n/g}$, $n\ge 1$  
($g<0$ always). 
The third and last consequence of eq.(\ref{heavylim}) is a severe limitation 
of perturbation theory: quantities involving in an essential way an infinite number 
of resonances cannot be described by means of a power series in $g$; the convergence 
radii $R_n$'s of the resonances involved indeed accumulate to zero in this case.
To excite the $l$-th resonance, one can use an initial function $\phi^{(l)}(x,t=0;\,g,N)$ 
in which the contributions of the resonances with order $n=1,2,\cdots, l-1,l+1,\cdots N$ 
have been subtracted, with $l\le N$.
Let us remark that $\phi^{(l)}(x,t=0;\,g,N)$ can also be computed in perturbation theory for 
$|g| \lsim 1/(2\pi N)$.

On the mathematical side, we have constructed the Riemann surface 
$S$ of the multivalued function $w=h(z)$ by gluing together the sheets $S_n$'s,
where the $h_n(z)$'s are defined.
We found that all the $S_n$'s with $n\ne 0$ are connected via the order-one branch 
points in $c_n$ and at $\infty$, to $S_0$.
All the resonance sheets $S_n$, $n\ne 0$, therefore are not
directly connected to each other, but only through the bound-state sheet $S_0$,
with the more general coupling one could think of, namely square-root 
branch points.
In physical terms, it is remarkable that the resonances 
"talk to each other" only indirectly, via the bound-state,
which does not even appear in the spectrum of the model
in the repulsive case, i.e. for $g>0$. 

\vskip 0.5truecm

\centerline{\bf Acknowledgments}

\vskip 0.5truecm

One of us (U.A.) wishes to thank Prof. G.~Parisi and Prof. M.~Testa 
for discussions.

\appendix

\section{The Lambert $W$ Function}
\label{appA}

The Lambert $W$ function is defined as the formal
inverse of the entire function
\beq
z = g(w) \equiv w \, e^w,
\eeq
i.e.  $w = W(z)$ (see \cite{knuth} for a detailed discussion). 
Symbolically:
\beq
W \, = \, g^{-1}.
\eeq
The function $g$ is not one-to-one because of the occurrence of
the (complex) exponential, so $W$ is multivalued.
The branch points of $W$ can be determined much in the same way
as we have made for the function $h$.
For example, since $g'(-1)=0$ while $g''(-1)\ne 0$, the Lambert
function has an order-one branch-point in $z=g(-1)=-1/e$.
$W$ also has infinite-order branch points in $z=0$ and $z=\infty$, 
so it can by considered as a generalization
of the complex logarithm. It is actually one of the simplest 
multivalued functions having both algebraic and logarithmic
branch points.
There is a branch of $W$ which is analytic in the origin,
where it vanishes, which is called the principal branch
and which we denote by $W_*$. 
The principal branch is real for a real argument $z$ in the 
range $-1/e \le z < \infty$ (see fig.\ref{figura_W}).
We call $W_0$ the branch connected to $W_*$ via the order-one 
branch point in $z=-1/e$. The zero branch is real for $-1/e\le z<0$.

\subsection{Connection to the $h$ Function}

Let us see how $W$ is related to our function $h$.
Let us assume $z \ne 0$ and introduce a variable $\xi=\xi(z)$
such that:
\beq
h(z) \, = \, - \, \frac{1}{z} \, - \, \xi \, .
\eeq
By applying $h^{-1} = f$ on both sides, we obtain:
\beq
\xi \, e^{ \xi} \, = \, - \frac{ e^{-1/z} }{z} \, .
\eeq
Since $W^{-1}(\xi) = g(\xi) = \xi \, e^{\xi}$,
the following formal relation holds between 
the Lambert function and the function $h$:
\beq
\label{relW*h0}
h(z) = \, - \, \frac{1}{z} \, - \, 
W \left( - \frac{ e^{-1/z} }{z} \right) .
\eeq
Let us stress however that the above equation involves
multivalued functions on both sides so, to have an
effective utility, one has to specify the branches
of both functions.
Let us also remark that the above equation, as it stands,
cannot be directly applied to study a neighborhood of $z=0$
of the branches $h_n$ with $n\ne 0$, as the latter are
analytic in the origin.

Let us see a simple consequence of the above relation.
Since the Lambert function $W(\zeta)$ has an order-one 
branch point in $\zeta=-1/e$, we derive that $h(z)$
has order-one branch points whenever
\beq
\frac{ e^{-1/z} }{z}  = \frac{ 1 }{ e },
\eeq
as we have already found directly (see eq.(\ref{solinz})).

We can also show that $z=0$ is a non-isolated singularity.
Since the function
\beq
q(z) \equiv \frac{e^{-1/z}}{z}
\eeq 
has an essential singularity at the origin, 
according to Picard's theorem
it basically assumes every complex value
in any punctured neighborhood of the origin, 
such as for example
\beq
\dot I_\varepsilon(0) = 
\big\{ z \in \CC : 0< |z| < \varepsilon \big\}, 
\, \, \, \, \, \, \varepsilon > 0. 
\eeq
One can consider paths $\gamma$ contained in $\dot I_\varepsilon(0)$
with images $q[\gamma]$ going around any number of times
the branch points of the Lambert $W$ function. 

\section{Evaluation of the Branch Points $c_n$'s of $h$}
\label{appB}

In this section we evaluate the (in general complex) zeroes of
the transcendental equation
\beq
\label{hard}
\frac{\exp(-1/z)}{z} = \frac{1}{e} ,
\eeq
which determine the order-one branch points of the function $h$,
with the exception of the point $z=1$ which, as shown in the
main text, is a zero but is not a branch point.

Let us first present a qualitative argument showing that
there cannot be any large $c_n$, i.e. any $c_n$ with $|c_n|\gg 1$, 
so that all the zeroes fall inside a limited region of the
$z$-plane.
For $|z|\gg 1$, one can expand the exponential around the
origin to give
\beq
\left| \frac{\exp(-1/z)}{z} \right|
\simeq \left| \frac{1}{z} - \frac{1}{z^2} + \frac{1}{2z^3} + \cdots \right| 
\ll 1 ,
\eeq
while the right-hand-side is of order one. Therefore we found
a contradiction by assuming $|c_n| \gg 1$.
The above argument can be refined by taking the imaginary
part on both sides of eq.(\ref{hard}):
\beq
{\rm Im} \left[ \frac{\exp(-1/z)}{z} \right] = \, 0 .
\eeq
By writing $z$ in polar coordinates,
\beq
z \, = \, \rho \, e^{i \varphi} \, ,
\eeq
with 
\beq
0 < \rho < \infty 
\, \, \, \, \, \, \, \, \, {\rm and} \, \, \, \, \,
-\pi < \varphi < \pi  ,
\eeq
the solution of the above equation is the union
of the sequence of curves 
\beq
\label{imagpart}
\alpha_n: \, \, \rho \, = \, \rho_n(\varphi) \, = \, 
\frac{\sin \varphi}{\varphi + n \pi} ,  
\eeq
where $n$ is any integer (see fig.\ref{figura_appB}).
Note that
\beq
\left| \rho_n(\varphi) \right| \le  \frac{1}{\pi\big(|n|-1\big)}
\to 0 \, \, \, \, \, \, \, {\rm for} \, \, \, n \to \pm \infty,  
\eeq
implying that the $\alpha_n$'s
shrink to zero in that limit.
By requiring the radius to be positive, 
we derive the following angular ranges
for the $\alpha_n$'s:
\bea
\rho_0&:& \left( -\pi, + \pi \right) ;
\\
\rho_n&:& \left[0, + \pi \right)  
\, \, \, \, \, \,\,\, {\rm for} \, \, n > 0 
\, \, \, {\rm and} \, \, \,  \left(-\pi, 0 \right] 
\, \, \, \, \, \,\,\, {\rm for} \, \, n < 0 .
\eea
Therefore $\alpha_n$ lies in the upper half-plane for $n>0$
and in the lower one for $n<0$.
Since the modulus of the r.h.s. of eq.(\ref{imagpart})
is $\le 1$, it follows that all the roots of (\ref{hard}) 
lie inside a closed disk of radius one centered at the origin.
Note that
\beq
\label{coincide}
z = z_n(\varphi) = 
\frac{ \exp(2i\varphi) - 1 }{ 2i\varphi + 2 \pi i n } ,
\eeq
i.e. $\alpha_n$ is the image of a segment of the imaginary
axis of $w$-plane.

We now prove that the branch-point equation (\ref{hard})
has two infinite sequences of solutions, $\{c_n\}_{n>0}$
and $\{c_n\}_{n<0}$, both converging to zero 
and contained in a curve with a  vertical tangent at the origin.
It follows in particular that 
\beq
\frac{ {\rm Re} \, c_n }{ {\rm Im} \, c_n} \to 0 
\, \, \, \, \, \, \, \, {\rm for} \, \, \, \, \, 
n \to \pm \infty .
\eeq
All the branch points lie on the curve, let's call it $\beta$, obtained 
by taking the modulus on both sides of eq.(\ref{hard}),
\beq
\label{fish}
\beta: \, \, \cos \varphi \, 
= \, - \, \rho \, \ln \left( \frac{\rho}{e} \right) 
\eeq
for $0 < \rho \le 1$.
As $\varphi=\varphi(\rho)$, let us also set for convenience 
$\varphi(0)=\pm \pi/2$, where the sign is determined by continuity.
The curve $\beta$ is symmetric for $\varphi\to-\varphi$,
has a vertical tangent in $z=0$, a cusp in $z=1$ and bounds 
a ``fish-like'' region (see figs.\ref{figura_appB} and 
\ref{figura5}).
The last condition to impose, to avoid spurious solutions, is
\beq
{\rm Re} \left[ \frac{\exp(-1/z)}{z}  \right] \, > \, 0 ,
\eeq
giving
\beq
2 \pi k  - \frac{\pi}{2}  <   \frac{\sin\varphi}{\rho} -\varphi 
<  2 \pi k  + \frac{\pi}{2} ,  
\eeq
for some integer $k$. 
By replacing eq.(\ref{imagpart}) in the above inequality,
we obtain
\beq
2 k  - \frac{1}{2} < n < 2 k + \frac{1}{2},
\eeq
implying that only even $n$'s are allowed:
\beq
n = 2 k
\eeq
with $k$ any integer .
By combining the above results, we obtain that
any branch point $c_k$ lies at the 
intersection of the curve $\alpha_{2k}$ (eq.(\ref{imagpart}))
with the curve $\beta$ (eq.(\ref{fish})):
\beq
c_k = \alpha_{2k} \cap \beta 
\, \, \, \, \, \, \, \, \, \, \, \, \, 
\forall k \ne 0 .
\eeq 
There is an infinite number of such intersections,
contained in two sequences converging at the origin
along $\beta$, as claimed.

\subsection{Explicit Solution}

Let us now derive an explicit approximation for $c_n$
for $n\gg 1$.
Since $\exp(-1/z)$ has an essential singularity in $z=0$,
it is rapidly varying close to the origin compared to $z$ so,
to a first approximation, we can fix $z$ to an arbitrary 
value, i.e. set $z=C \ne 0$ (the exponential never vanishes). 
We can also assume $C=\mathcal{O}(1)$,
as we have already proved that there are no large solutions.
The dependence on $C$ actually cancels order by order of the 
iterations of the equation derived later in this section.
The original equation then simplifies to
\beq
\exp(-1/z) \, = \, \frac{C}{e} ,
\eeq
having the solutions
\beq
z_n = \frac{i}{2\pi n + i \ln e/C},
\eeq
with $n$ any nonzero integer. By writing 
\beq
c_n = \frac{ i }{ 2 \pi n + i \ln e/C + k_n} , 
\eeq
one obtains the (still exact) equation for $k_n$
\beq
e^{i k_n} = \frac{ i }{ C ( 2 \pi n + i \ln e/C + k_n ) } , 
\eeq
which can be used to set up a recursion scheme
generating a series converging to $c_n$.
By writing
\beq
k_n = \sum_{l=1}^\infty k_n^{(l)} = k_n^{(1)} + k_n^{(2)} + k_n^{(3)} + \cdots
\eeq
and assuming (as is verified a posteriori)
\beq
|n| \gg \left|k_n^{(1)}\right| \gg \left|k_n^{(2)}\right| \gg \left|k_n^{(3)}\right| \gg \cdots
\eeq
for $|n|\gg 1$, one obtain for example at first order:
\beq
e^{i k_n^{(1)}} = \frac{ i }{ C ( 2 \pi n + i \ln e/C )  } .
\eeq
The second recursion produces the following formula, which is
quite accurate and is used in the main text (in which we have
set at the end $C=e$):
\beq
c_n \, = \, 
\frac{i}
{ \pi(2n+1/2) + i\ln\big[ e \pi (2n+1/2) \big] 
- \ln\big[ e \pi (2n+1/2) \big]/(\pi(2n+1/2))}
+ \mathcal{O}\left[ \frac{ 1 }{ ( 2 \pi n )^4 }  \right] ,
\eeq
for $n\ge 1$. For $n<0$, one just uses the symmetry relation in 
eq.(\ref{simmetria}).
Note that the expansion parameter is actually $1/(2\pi n)$, so there
is a rather good convergence of the above series even for $n=1$.
The images of the branch points read:
\beq
d_n = i \pi\left(2n+\frac{ 1 }{2 }\right) 
- \ln\left[ \pi \left(2n+\frac{ 1 }{2 }\right) \right] 
- i \, \, \frac{  \ln\big[ e \pi (2n+1/2) \big]   }{  \pi(2n+1/2)    }
+ \mathcal{O}\left[ \frac{ 1 }{ ( 2 \pi n )^2 }  \right] .
\eeq

To summarize, we have obtained two sequences of branch points of
order one for $h$ for $n>0$ and $n<0$, both converging to zero for 
$n \to \pm \infty$ respectively.

\section{Absence of Spurious Branches in Asymptotic Expansions}
\label{appC}

Since the function $w=h(z)$ has a logarithmic branch point in $z=\infty$,
one expects that the expansion around infinity involves 
powers of $\ln(z)$ only, plus of course eventual single-valued 
functions of $z$.\footnote{
That is the generalization to logarithmic branch points
of the Puiseaux expansion around a branch point
$a$ of order $n-1$, which involves the integer powers of $(z-a)^{1/n}$
\cite{donna};
actually, it is the formal limit for $n\to\infty$.}
By looking at the asymptotic expansion obtained for $h(z)$,
the above fact literally is not true, as the composition
of the complex logarithm with itself does appear.
It may seem therefore that the expansion produces spurious 
multivaluedness.
In  this appendix we show that actually this is not the case,
namely that the "external logarithm'' does not give rise to 
any additional (and undesired) multivaluedness.

Let us sketch the proof for the simplest non-trivial case,
the expansion to first order (the first two terms).
Our expansion,
\beq
h_n(z) = \ln_n(z) + \ln_0\left[\ln_n(z)\right] 
+ \mathcal{O}\left[\frac{ 1 }{ \ln_n(z)  }\right] ,
\eeq
involving the principal branch of the external logarithm,
has to be compared with an expansion involving a different
branch of the external logarithm:
\beq
\tilde{h}_n(z) = \ln_m(z) + \ln_k\left[\ln_m(z)\right] 
+ \mathcal{O}\left[\frac{ 1 }{ \ln_m(z)  }\right] ,
\eeq
where $k$ is a fixed integer, while $|n| \gg 1$ by assumption. 
By setting
\beq
n = m + k ,
\eeq
the difference between our formula and the new one
reads:
\beq
h_n(z) - \tilde{h}_n(z) = 
\log_0\left[ 1 + \frac{ 1 }{ \ln_0(z) + 2\pi i (n-k) } \right] 
+ \mathcal{O}\left(\frac{ 1 }{ n }\right) 
= \mathcal{O}\left(\frac{ 1 }{ n }\right) ,
\eeq
i.e. the difference of the logs is absorbed by the higher
order terms, as we claimed. 

Let us note that the two expansions actually produce slightly different
numerical values.
In general, there is an ambiguity in such truncated expansions,
which is never removed, but only pushed to higher orders by
adding more and more terms. 

\end{document}